\documentclass[prd,twocolumn,aps,superscriptaddress,amsmath,amssymb,amsfonts,nofootinbib]{revtex4-2}

\pdfoutput=1

\usepackage{graphicx}
\usepackage{xcolor}
\usepackage{slashed}
\usepackage{tabularx}
\usepackage{hyperref}
\usepackage{bbold}
\usepackage{bbm}
\usepackage{amsmath}
\usepackage{placeins}
\usepackage[normalem]{ulem}
\usepackage{dsfont}
\usepackage{textcomp}
\usepackage{gensymb}
\usepackage{cleveref}
\usepackage{comment}

\newcommand{\esig}{\ensuremath{\langle\sigma\rangle}}
\newcommand{\ourg}{\tilde g}
\newcommand{\mbare}{{\mathring m}_\sigma}

\begin{document}

\newcommand{\JLU}{Institut f\"ur Theoretische Physik, Justus-Liebig-Universit\"at, Heinrich-Buff-Ring 16, 35392 Giessen, Germany}
\newcommand{\HFHF}{Helmholtz Forschungsakademie Hessen f\"ur FAIR (HFHF), GSI Helmholtzzentrum f\"ur Schwerionenforschung, Campus Giessen}
\newcommand{\UU}{Institutionen f\"or fysik och astronomi, Uppsala universitet, Box 516, S-75120 Uppsala, Sweden}
\newcommand{\UCY}{Department of Physics, University of Cyprus, P.O. Box 20537, 1678 Nicosia, Cyprus}

\title{Kinetic Mixing and Axial Charges in the Parity-Doublet Model}

\author{Christian Kummer}\affiliation{\JLU}\affiliation{\UCY}
\author{Stefan Leupold}\affiliation{\UU}
\author{Lorenz von Smekal}\affiliation{\JLU}\affiliation{\HFHF}

\date{\today}

\begin{abstract}
The standard parity doublet model with its mass-mixing mechanism fails to describe the axial charge $g_A$ of the nucleon. While $g_A = 1$ in the original Gell-Mann--Levy model, which reproduces the Adler-Bell-Jackiw anomaly of QCD,  in the presence of a chirally invariant baryon mass the mass mixing leads to $g_A < 1 $ whereas phenomenologically it is about 1.28. We propose to remedy this problem by introducing kinetic-mixing terms corresponding to meson-baryon derivative couplings, similar in spirit to the two-mixing-angle scenario of the $\eta$-$\eta'$ mixing. This extended parity doublet model contains five parameters in the effective baryonic Lagrangian. Three of them can be determined by using the empirical results for the axial charge of the nucleon together with the masses of the nucleon and its parity partner, the $N^*(1535)$ resonance. We discuss various options how to determine the remaining parameters, touching upon the mass of both parity partners if the chiral condensate is put to zero; the mass of the nucleon in the chiral limit; and the values of meson-baryon coupling constants related to the decays of the resonance to pion-nucleon and sigma-nucleon.
\end{abstract}

\maketitle

\section{Introduction}

The minor part of the mass of the visible universe is dynamically generated by the Higgs mechanism. The major part is dynamically generated by QCD \cite{Wilczek:1999be}. An interesting aspect about dynamical mass generation is the fact that the mechanism can be quantitatively influenced by changing from the vacuum to a medium with finite energy density and/or finite net particle densities. In thermal equilibrium, those systems might be characterized by finite temperature and/or chemical potential(s). For the Higgs mechanism, one needs huge energy densities/temperatures on the scale of electroweak symmetry breaking \cite{Kapusta:2006pm}. This relates to the physics of the early universe. But for QCD, one needs ``only'' energy densities that can be reached in laboratory experiments with relativistic nucleus-nucleus collisions \cite{Friman:2011zz}. In addition, already the particle density inside of atomic nuclei might be large enough to cause changes in the masses of hadrons \cite{Leupold:2009kz}. 

One interesting question is the relation between mass generation and chiral symmetry breaking \cite{Nambu:1961tp,Nambu:1961fr,Ioffe:1981kw,Detar:1988kn,Brown:1991kk,Jido:2001nt,Rapp:2009yu,Glozman:2022zpy,Hilger:2010cn,Weyrich:2015hha,Larionov:2021ycq,Kim:2021xyp}. In a world without spontaneous chiral symmetry breaking, there would be nearly degenerate chiral partners of opposite parity, i.e.\ with mass differences on the order of isospin-breaking effects, like the mass difference between proton and neutron. Clearly, nature (the vacuum) is not like that. The mass differences between hadrons of opposite parity are comparable to the hadron masses themselves \cite{ParticleDataGroup:2024cfk}. Thus, the quark condensate has a large influence on the spectrum of hadrons. Of course, the values of condensates are going to change in a medium. If all the mass of say the nucleon, and hence the visible universe, was generated by the effect of spontaneous chiral symmetry breaking, then the nucleon could be nearly massless in a dense enough environment leading to an abnormal state of Lee-Wick matter \cite{Lee:1974ma}. 

On the other hand, the traditional success of QCD sum rules to relate the masses of charmonium states to the gluon condensate \cite{Shifman:1978by} suggests that it is not only the quark condensate that influences the hadron masses. Yet, one should not view this as two entirely unrelated mechanisms. There are gluon configurations that cause chiral symmetry breaking and therefore a non-vanishing quark condensate. However, there are also gluon configurations that can cause a non-vanishing gluon condensate for systems that are so dense that chiral symmetry is restored. 

If one wants to study how the properties of hadrons are modified when changing from vacuum to a dense, strongly interacting medium, then such a formalism must naturally satisfy two requirements: first, a good account of the physics of the starting point, the vacuum; second, a non-perturbative quantum field theory that can reliably deal with the strong many-body aspects of the medium. The present paper has a focus on the first aspect, the vacuum situation. But we will formulate the framework such that it can be applied for in-medium situations. 

The Gell-Mann--Levy model \cite{Gell-Mann:1960mvl} highlights the impact of spontaneous chiral symmetry breaking on the mass of the nucleon. Here, the entire mass of the nucleon is caused by a non-vanishing expectation value of the field of the sigma meson. In view of the previous discussion about the importance of a gluonic contribution that does not vanish at chiral restoration, the Gell-Mann--Levy model is too simple. 
In addition, the aspect of chiral partners with opposite parity appears only for the pion and sigma meson, but the model does not include a parity partner for the nucleon. Concerning its vacuum phenomenology, the axial charge $g_A$ of the nucleon is exactly $1$ which, on the one hand, matches the Adler-Bell-Jackiw triangle anomaly \cite{Adler:1969gk,Bell:1969ts} for $\pi_0 \to 2\gamma $ \cite{Steinberger:1949wx,Bar:2001qk} but, on the other, requires radiative corrections to agree with measurements of the beta decay of the neutron yielding a value of $ g_A \approx 1.28$ \cite{ParticleDataGroup:2024cfk}.

The parity doublet model (PDM) \cite{Detar:1988kn,Jido:2001nt} is an extension of the Gell-Mann--Levy model that allows a distinction between a mass contribution caused by chiral symmetry breaking and a purely gluonic contribution that is formally untouched by chiral restoration. It also includes the nucleon $N$ together with a parity partner. For the present work we assume that this partner is the $N^*(1535)$ resonance. The original two baryon states have opposite parity, are degenerate at chiral restoration, but transform oppositely with respect to chiral transformations and hence have different couplings to the sigma-pion multiplet. When chiral symmetry is spontaneously broken, the two baryon states mix and form the non-degenerate states of $N$ and $N^*(1535)$. The mixing angle $\theta $ can be related to the original parameters of the model which reduces to a Gell-Mann--Levy model with two independent isospin doublets $N$ and $N^*$ for $\theta = 0$ without mixing. As we will review below, in presence of the mass mixing of the standard PDM, the axial charges $g_A$ and $g_A^*$ of both baryons remain equal and are given by $g_A = g_A^* = \cos 2\theta $ \cite{Jido:2001nt}.
Although the PDM allows for appealing in-medium studies, this particular aspect of vacuum phenomenology is thus worse than in the original Gell-Mann--Levy model. In fact, the larger the fraction of the chirally-invariant gluonic mass of the baryons is, the smaller are their axial charges in the standard PDM.

In general, the way how to overcome this problem is clear; one needs extra terms in the Lagrangian. The problem is not to invent new terms. The problem is then rather to argue why to include a specific term and not another one. In view of the Lagrangian of the original version of the model with baryon terms that have either none or one derivative, we propose here to keep this restriction and abandon higher-derivative terms. The strategy is to regard the model as a low-energy proxy to QCD, for which it certainly makes sense to ignore interactions with too many derivatives. In doing so, we apply the spirit of effective field theories and include all terms allowed by symmetry with either zero or one derivative in the baryon sector. We also will not follow the suggestion to add axial-vector fields to the theory and mix their longitudinal components with the pion fields \cite{Gallas:2013ipa}, because such a 
mixing is absent when describing (axial-)vector mesons as resonance poles in conserved current correlators by anti-symmetric rank-two tensors in the Joos-Weinberg representation for massive spin-1 fields  \cite{Jung:2019nnr,Tripolt:2021jtp}.    

Inclusion of extra terms with one derivative implies a modification of the kinetic terms of the baryons. In the most general case (considered here) this leads to an additional mixing between the parity partners. One mixing appears now in the ``mass'' sector, the terms without derivatives, and one in the kinetic sector. The corresponding mixing scenario can be formulated with two mixing angles. It resembles the situation in the boson sector, discussed in the literature for instance for the mixing of the two pseudo-scalar mesons $\eta$ and $\eta'$, which leads to a significantly improved phenomenology \cite{Escribano:2005qq}. For renormalization aspects in the context of kinetic mixing see \cite{Bijnens:2018rqw}. 

The paper is structured in the following way. To set the stage, we discuss the standard version of the PDM in section \ref{sec:standard}. In Section \ref{sec:KineticMixing}, we present our extended version of the model. Two mixing angles are introduced to decouple the baryon fields. 
We relate the model parameters to the mixing angles and baryon masses.
The formulae derived in Subsection \ref{subsec:KML} are manifestly Lorentz invariant. For a discussion of vacuum properties this formalism is sufficient. However, it is illuminating to discuss the mixing scenario also from the point of view of a Dirac Hamiltonian instead of just a Lagrangian. We present this alternative derivation in Subsection \ref{sec:NGDH}. With this subsection we prepare the ground for a future generalization to an in-medium situation. 
In Section \ref{sec:baryon-meson-couplings} we make contact between the PDM parameters and the three-point interactions that couple mesons to baryons. In addition, we cross-check the extended Goldberger-Treiman relations. In Section \ref{sec:results}, we discuss the determination of the parameters of our model. In the standard PDM one has three parameters. They cannot be determined by the masses of the two baryons alone. One parameter remains undetermined, i.e.\ must be taken from some other source of information. One might use the mass $m_0$ of the parity partners in the absence of chiral symmetry breaking as this undetermined parameter. Correspondingly, we have five parameters in our extended model but only four clear-cut phenomenological values: the two baryon masses, the axial charge of the nucleon, and the coupling of the $N^*(1535)$ to the pion-nucleon system. We explore the consequences of reasonable choices of the mass parameter $m_0$. We summarize our results in Section \ref{sec:summary}. Appendices are added to cover additional aspects that would interrupt the flow of the main text. 

\section{Standard Version of the Parity Doublet Model}
\label{sec:standard}

The PDM provides an effective hadronic description of chiral symmetry breaking in the presence of a chirally invariant baryon mass $m_0$, common to nucleons $N$ and the negative-parity baryon doublet $N^*(1535)$ as their natural parity partners. This is possible with the so-called mirror assignment of two sets of spin-$1/2$ baryons in two-flavor chiral $(\frac{1}{2},0) \oplus (0, \frac{1}{2}) $ representations $N_1$ and $N_2$ \cite{Detar:1988kn,Jido:2001nt}. The baryon-meson interactions are chirally invariant, but spontaneous chiral symmetry breaking introduces additional Dirac mass terms $m_1 $ and $m_2 $ in the chiral representations and hence a non-trivial mixing between those and the mass eigenstates of $N$ and $N^*$. In a more modern language, the rationale for such a contribution to the mass of the visible universe are the so-called gravitational form factors of the nucleon, i.e.~the matrix elements of the QCD energy-momentum tensor between nucleon states, whose purely gluonic contributions from the QCD scale anomaly do not rely on chiral symmetry breaking \cite{Yang:2018nqn,Burkert:2023wzr}.

In terms of the chiral representations $N_1$ and $N_2$, the baryon Lagrangian of the PDM with its $O(4)$-invariant Yukawa couplings to the scalar $\sigma$-meson and the pseudo-scalar pions, analogous to those in the Gell-Mann--Levy model, and an off-diagonal mass term proportional to $m_0$ is given by  
\begin{align}
  \mathcal{L} \, =& \,  \overline N_1 \big[i\slashed{\partial} - g_1 (\sigma + i\gamma_5    \vec{\tau}\vec{\pi})\big] N_1  
  \nonumber \\  &  \hskip .4cm
  + \overline N_2 \big[i\slashed{\partial} - g_2 (\sigma - i\gamma_5 \vec{\tau}\vec{\pi})\big] N_2 
  \label{eq:chiralLagr} \\ 
  &  \hskip .8cm
  - m_0 \big(\overline N_1 \gamma_5 N_2 - \overline N_2 \gamma_5 N_1\big)  \,.      \nonumber
\end{align}
The relative minus sign in the pion-baryon couplings reflects the mirror assignment. 
In the presence of spontaneous symmetry breaking, the $\sigma$ field receives a non-vanishing vacuum expectation value 
$\langle \sigma \rangle = f_\pi \approx 92 \,$MeV, leading to Dirac mass contributions $m_1= g_1 \langle\sigma\rangle$  and $m_2= g_2 \langle\sigma\rangle$  to $N_1$ and $N_2$.
The physical positive $N_+=N$ and negative $N_- = N^*$ parity baryon fields are obtained from diagonalizing the mass matrix in the Nambu-Gorkov space of parity partner baryons using a simple rotation, with 
\begin{equation}
   \begin{pmatrix}
          N_+\\
          N_-
   \end{pmatrix}
 = \begin{pmatrix}
          \cos\theta         &  \gamma_5\sin\theta \\
         -\gamma_5\sin\theta &  \cos\theta
    \end{pmatrix}
   \begin{pmatrix}
          N_1\\
          N_2
   \end{pmatrix}\, . \label{eq:PhysFields}
\end{equation}
The mixing angle $\theta$ is obtained from the condition that the off-diagonal elements contain only interaction terms between baryons, pions and the  fluctuating $\sigma$-meson field $\tilde \sigma = \sigma - \langle\sigma\rangle$. It turns out to simply be determined by the ratio of the chirally invariant (gluonic mass)  $m_0$ and the average mass from chiral symmetry breaking, $\overline m_\chi = (m_1 + m_2)/2$,
\begin{equation}
    \tan 2\theta = \frac{m_0}{\overline{m}_\chi} \, . \label{eq:theta_mix}
\end{equation}
The Lagrangian in Eq.~\eqref{eq:chiralLagr} then becomes
\begin{align}
  \mathcal{L} =& \, \overline N_+ 
  \left( i\slashed{\partial} - m_+  \right) N_+ \\
  &\hskip .4cm + \overline N_- \left( i\slashed{\partial} - m_-  \right)
  N_- +   \mathcal{L}_\mathrm{int} \, ,\nonumber
  \end{align}
where the physical masses $m_+ = m_N$ and $m_- = m_{N^*}$ of the positive and negative parity baryons are given by
\begin{align}
  m_+ =& \,  m_0 \sin 2\theta 
  + 
  m_1 \cos^2\theta - 
  m_2 \sin^2\theta \,, \nonumber \\
  m_- =& \, m_0 \sin 2\theta 
  + 
  m_2 \cos^2\theta - 
  m_1 \sin^2\theta  \, . 
  \label{eq:mass_pm}  
\end{align}
These masses will later frequently also be expressed as $m_\pm = \overline m \pm \Delta m $, 
in terms of their average $\overline m = (m_++m_-)/2$ and $\Delta m = (m_+-m_-)/2$, with
\begin{align}
  \overline m 
  =& \, m_0 \sin 2\theta + 
 \overline m_\chi  \cos2\theta \,, \;\;
  \Delta m  
  = \, (m_1- m_2)/2\, . 
  \label{eq:mass_ad}  
\end{align}
For $m_0=0 \Leftrightarrow \theta = 0 $ (restricting $\theta $ to the principal branch of the arctangent), this reduces to  $m_+ = m_1$ and $m_- = m_2$, and for $\esig = 0 \Leftrightarrow \theta = \pi/4 $ (restricting to positive masses $ m_0$, $\overline m_\chi \ge 0 $), we recover $m_+=m_-=m_0 $.  On the other hand, with positive $\esig = f_\pi $, and $m_1$, $m_2 \ge 0$, but for $\Delta m = (m_N - m_{N^*})/2 < 0 $ this immediately implies that we are looking for solutions with $0 < g_1 < g_2\,$. In fact, the second relation in \eqref{eq:mass_ad} implies, in the vacuum, 
\begin{equation}
g_2-g_1 =     \frac{m_{N^*}-m_N}{f_\pi} \approx 6.21 \,, 
\end{equation}
with $m_N= 939\,$MeV, $m_{N^*} = 1510\,$MeV, and $f_\pi = 92\,$MeV \cite{ParticleDataGroup:2024cfk}. Two important auxiliary relations for later, which are easily derived from Eqs.~\eqref{eq:theta_mix}, \eqref{eq:mass_pm} and \eqref{eq:mass_ad}, are
\begin{align}
      m_1 \cos^2\theta + m_2\sin^2 \theta =&\,  m_+ \cos2\theta   \,, \nonumber\\
      m_2 \cos^2\theta + m_1\sin^2 \theta =&\,  m_-  \cos2\theta  \, .\label{eq:GT_aux}
\end{align}

Before we move on with our discussion of this standard version of the PDM, we would like to point out that certain parameter choices would lead to a different physical situation in the Nambu-Goldstone phase of spontaneous chiral symmetry breaking. Actually one can obtain two non-degenerate baryons that have both the {\em same} parity instead of opposite parities. This case is further discussed in Appendix \ref{sec:flip}. We do not regard it as likely that this situation is realized in nature. Still we find it interesting enough to discuss this possibility. In the main text we will stick to the standard scenario of two baryons with opposite parity.

The interaction terms of the baryons and the fluctuating (pseudo-)scalar meson fields $\tilde\sigma $ and $\pi$ are best collected in a matrix notation. For later convenience we first represent the $O(4)$-vector of $(\tilde\sigma,\vec \pi) $ by a meson matrix $\tilde M$ which is quaternion-block diagonal in the chiral representation,
\begin{equation}
    \tilde M = \tilde\sigma + i\gamma_5    \vec{\tau}\vec{\pi} \, .
\end{equation}
Furthermore, using the abbreviations $c=\cos\theta$ and $s=\sin\theta $, the interaction Lagrangian can then be written as
\begin{widetext}
\begin{align}
    \mathcal{L}_\mathrm{int} &= -
    \begin{pmatrix}
        \overline N_+ ,  \overline N_-
    \end{pmatrix}  
    \begin{pmatrix}
        c^2 \, g_1 \tilde M - s^2  \, g_2 \tilde{M}^\dagger &
        -\gamma_5 \, cs \, \big(g_1 \tilde M + g_2 \tilde M^\dagger \big) \\[2pt] 
       \gamma_5 \, cs \, \big(g_1 \tilde M + g_2 \tilde M^\dagger \big) & - s^2 \, g_1 \tilde M +  c^2  \, g_2 \tilde M^\dagger
    \end{pmatrix}
    \begin{pmatrix}
          N_+\\
          N_-
   \end{pmatrix} \, .
\end{align}
This defines the corresponding physical baryon-meson couplings,
\begin{align}
    \mathcal{L}_\mathrm{int} &=
    - \begin{pmatrix}
        \overline N_+ ,  \overline N_-
    \end{pmatrix}  
    \begin{pmatrix}
        g_{\sigma NN} \, \tilde\sigma + g_{\pi NN} \, i\gamma_5  \vec{\tau}\vec{\pi} &  - g_{\sigma N N^*}\, \gamma_5 \tilde\sigma + g_{\pi NN^*} \, i  \vec{\tau}\vec{\pi} \\[2pt]
        g_{\sigma N N^*} \, \gamma_5 \tilde\sigma - 
        g_{\pi NN^*} \, i  \vec{\tau}\vec{\pi} & 
        g_{\sigma N^*N^*} \tilde\sigma -g_{\pi N^*N^*} \, i\gamma_5  \vec{\tau}\vec{\pi}
    \end{pmatrix}
    \begin{pmatrix}
          N_+\\
          N_-
   \end{pmatrix} \, .
\end{align}
\end{widetext}
Note that the signs of the couplings on the diagonal have been chosen to correspond to those in the original Lagrangian in \eqref{eq:chiralLagr}, without mixing, whereas the off-diagonal interactions have been chosen arbitrarily here to fix our notations. We can then read off the baryon-meson couplings by comparison. For the $\sigma $ resonance this yields,
\begin{align}
    g_{\sigma NN} &=\,  g_1 \cos^2\theta - g_2 \sin^2\theta \, ,\nonumber \\
    g_{\sigma NN^*} &=\, (g_1+g_2) \sin\theta\cos\theta \, , \label{eqs:sigBB} \\
    g_{\sigma N^*N^*} &=\, g_2 \cos^2\theta - g_1 \sin^2\theta \, , \nonumber
\end{align}
and for the pion,
\begin{align}
    g_{\pi NN} &=\,  g_1 \cos^2\theta + g_2 \sin^2\theta \, ,\nonumber \\
    g_{\pi NN^*} &=\, (g_2-g_1) \sin\theta\cos\theta \, ,  \\
    g_{\pi N^*N^*} &=\, g_2 \cos^2\theta + g_1 \sin^2\theta \, . \nonumber
\end{align}
With the auxiliary relations in \eqref{eq:GT_aux}, we can express the latter, with $\esig =f_\pi $ in the vacuum, in the form, 
\begin{align}
    f_\pi \, g_{\pi NN} &=\,  m_N \cos 2\theta \, ,\nonumber \\
    f_\pi\,  g_{\pi NN^*} &=\, -\Delta m  \sin2\theta\, ,  \\
    f_\pi\, g_{\pi N^*N^*} &=\, m_{N^*} \cos 2\theta  \, . \nonumber
\end{align}
So far, the same was essentially derived already in Ref.~\cite{Jido:2001nt}. From the Goldberger-Treiman relations \cite{Goldberger:1958tr,Goldberger:1958vp}, $ m_N \, g_A = f_\pi\, g_{\pi NN} $ and $ m_{N^*} \, g_A^* = f_\pi\, g_{\pi N^*N^*} $, one then concludes for the axial charges $g_A \equiv g_A^{++} $ and $g_A^* \equiv g_A^{--} $ of the nucleon and the $N^*(1535)$, respectively, in the standard PDM,
\begin{equation}
    g_A = g_A^* = \cos2\theta \, ,  \label{eq:sPDMgA}
\end{equation}
and one can analogously define the transition coupling 
\begin{equation}
    g_A^{+-} = \frac{f_\pi \, g_{\pi NN^*}}{m_{N^*} - m_N } = \sin 2\theta\, . \label{eq:sPDMgApm}
\end{equation}

This has several problems. First of all, from neutron and nuclear beta decays the experimental value of the axial charge of the nucleon is  $g_A \simeq 1.28$ \cite{Falkowski:2020pma,ParticleDataGroup:2024cfk}. In fact, including the Goldberger-Treiman discrepancy, here one should probably rather use an even larger value of $g_A \simeq 1.3$, from a corresponding (isospin-averaged) pion-nucleon coupling $g_{\pi NN} \simeq 13.3$; for a recent determination of pion-nucleon couplings from effective field theory, see \cite{Reinert:2020mcu}. The standard PDM result with $g_A < 1 $ in \eqref{eq:sPDMgA} is certainly phenomenologically unacceptable.
The good old result $g_A=1$ of the Gell-Mann--Levy model, which matches the Adler-Bell-Jackiw (ABJ) triangle anomaly of QCD for $\pi_0 \to 2\gamma $ in this case, here only follows for $m_0 = 0$, i.e.\ without any contribution to the nucleon mass from the gluonic parts of its gravitational form factors.  Because the rationale for the PDM as an effective hadronic theory for chiral symmetry breaking is to describe just that, this is very unsatisfying,  especially when it comes to including weak interactions. 

Secondly, the axial charge $g_A^*$ of the $N^*(1535) $ is not at all equal or even comparable in size to $g_A $, as the standard PDM predicts at this level from \eqref{eq:sPDMgA} as well. By all evidence, from what we know, one rather has $g_A^* \ll g_A$ \cite{Takahashi:2008fy,An:2008tz}.  
In the standard PDM, on the other hand, they
result to be exactly the same, $g_A^* = g_A$, so that the respective triangle contributions to the ABJ anomaly of proton and $N^*$ exactly cancel each other, due to the mirror assignment being proportional to $ g_A - g_A^* $ with our sign conventions, here. 
With the phenomenological observation that $g_A^* \ll g_A $, where $g_A^*$ may even still be consistent with zero within errors \cite{Takahashi:2008fy}, the solution to this problem is not as simple as to add a Wess-Zumino-Witten term in order to account for the ABJ anomaly (as proposed at the time in Ref.~\cite{Jido:2001nt}). Instead of their perfect cancellation, within the present accuracy of $g_A^*$ it is phenomenologically quite viable that the sum of the parity-partner baryon triangles
does in fact match the ABJ anomaly of QCD with $g_A-g_A^* = 1 $. We will see below that with kinetic mixing in the extended PDM it is now possible, in principle, to have a sizable chirally invariant baryon mass $m_0$, and hence a sizable gluonic contribution to the mass of the visible universe, together with $g_A = 1.28 $ and $g_A^*$ small at the same time, e.g., also with matching the ABJ anomaly of QCD as in the Gell-Mann--Levy model, but here with $g_A-g_A^* = 1 $.

We close this section with a brief remark on the expected size of the chirally invariant nucleon mass $m_0$. A naive estimate could be obtained from the scalar gravitational form factor which is defined as the matrix elements of the trace of the energy-momentum tensor, measuring the non-conservation of the QCD dilatation current, in nucleon states at vanishing momentum transfer \cite{Yang:2018nqn,Burkert:2023wzr}. This leads to a nucleon-mass decomposition $m_N = m_m + m_A$, 
where  $m_m$ contains the classical contribution
that vanishes along with the nucleon sigma term in the chiral limit. 
If the gluonic contribution in $m_A$ was totally unaffected by chiral symmetry breaking, this would imply that $m_0 $ agrees with the nucleon mass in the chiral limit, which is about $880\,$MeV \cite{Owa:2023tbk,Procura:2003ig,Bernard:2003rp}. Because this is a drastic assumption, as the nucleon states are certainly expected to react to chiral symmetry breaking, it can at best serve as an upper bound.  As another estimate, according to the QCD sum rule analysis of Ref.~\cite{Kim:2021xyp}, the masses of nucleons and $N^*(1535)$ should converge to a common mass value of $m_0 $ between $500\,$MeV and $ 550\,$MeV in the chirally restored vacuum.

\section{Kinetic mixing}
\label{sec:KineticMixing}

\subsection{Effective Lagrangian with derivative couplings}
\label{subsec:KML}

A way to remedy the short-coming of $\vert g_A \vert \le 1$ is the inclusion of kinetic mixing instead of pure mass mixing. 
Kinetic mixing brings along derivative couplings. Besides the pseudo-scalar pion-nucleon interaction, which is standard for linear sigma models, one obtains also the well-known pseudo-vector interaction \cite{Ericson:1988gk,Scherer:2012xha}. For the interactions of pions, nucleons and baryon resonances one obtains even derivative couplings that cannot immediately be rewritten into pseudo-scalar or standard pseudo-vector type. Of course, by using free equations of motion for the baryon fields, i.e.\ by dropping interaction terms beyond three-point interactions, one is able to relate all three-point interactions to each other. 

In addition, kinetic mixing brings along modifications of the interaction of nucleons and left-handed currents. This is, of course, the desired aspect that allows for a modification of the axial charge. We will see that $\vert g_A \vert >1$ becomes possible. Given all these aspects, we have dedicated Appendix \ref{sec:details} to a detailed motivation and derivation of our extended PDM. 

The final result of our considerations is the following Lagrangian with 5 parameters:
\begin{align}
  \mathcal{L}_\mathrm{ext} =& \,   \overline N_1 \big( i\slashed{\partial} - g_1 
  M\big)  N_1  
  + \overline N_2 \big( i\slashed{\partial} - g_2 
  M^\dagger \big) N_2  
  \nonumber \\ 
& {}
  - m_0 (\overline N_1 \gamma_5 N_2 - \overline N_2 \gamma_5 N_1) \label{eq:Lagr-ext}\\ 
  & {} + h_1 \big( \overline N_1 \, 
  M \gamma_5 i \slashed{\partial} N_2 +
    \overline N_2 i \!\overleftarrow{\slashed{\partial}} \, 
    \gamma_5 M N_1 \big) 
    \nonumber \\ & {}
    + h_2 \big( \overline N_1 i \!\overleftarrow{\slashed{\partial}} \, 
  \gamma_5 M^\dagger N_2 + 
    \overline N_2 \, 
    \gamma_5 M^\dagger i \slashed{\partial} N_1 \big) 
  \,.      \nonumber
\end{align}
Here we have introduced the full meson field matrix $M =  \sigma + i\gamma_5    \vec{\tau}\vec{\pi}$, including the vacuum expectation value.\footnote{Note that in Appendix \ref{sec:details} we use a slightly different meson matrix together with the left- and right-handed fermion fields.} 

The terms $\sim h_1, h_2$ have been constructed such that they are manifestly hermitian. 
To make contact with standard kinetic 
terms we pull all derivatives to the right by partial integration (and compensate the sign change by anti-commuting $i\slashed{\partial}$ and $\gamma_5$). If we furthermore  split the couplings $h_{1,2}$ into sums and differences of the corresponding interaction terms, according to $h_{1,2}\equiv h \pm \Delta h$, 
 the extended PDM Lagrangian with kinetic mixing becomes
\begin{equation}
    \mathcal{L}_\mathrm{ext} = \mathcal L + \mathcal L^h + \mathcal L^{\Delta h}
\end{equation} 
where $\mathcal L$ is the standard PDM Lagrangian in \eqref{eq:chiralLagr}, coinciding with the $h_1$ and $h_2$ independent terms in \eqref{eq:Lagr-ext}. The kinetic mixing terms are collected in
\begin{align}
  \mathcal{L}^h =&  \,  h \big( \overline N_1 \gamma_5 \,  2M \, i \slashed{\partial} N_2   + \overline N_1 \gamma_5 (i \slashed{\partial}M^\dagger)  N_2 \nonumber\\
  & \hskip .6cm  +  \overline N_2 \gamma_5 \, 2 M^\dagger \,  i \slashed{\partial} N_1 
  + \overline N_2 \gamma_5 (i \slashed{\partial} M ) N_1 \big)  \, ,
     \label{eq:extLagr} \\ 
  \mathcal{L}^{\Delta h} =& \,  \Delta h \left(  \overline N_2 \gamma_5 (i \slashed{\partial} M ) N_1 
  - \overline N_1 \gamma_5 (i \slashed{\partial} M^\dagger)  N_2 \right)
  \,.       \nonumber
\end{align}
In this form, one observes that the interaction terms in $\mathcal L^{\Delta h} $ do not affect the mixing of the baryon fields for spacetime independent meson field expectation values, but only introduce additional (pseudo-)vector couplings. 

For the mixing of the chiral mirror representations $N_1 $ and $N_2$ in the vacuum, we can thus focus on $\mathcal L + \mathcal L^h $ and  use $M = M^\dagger = \esig = f_\pi $. This leads to the baryonic Lagrangian in the vacuum 
\begin{align}
    \mathcal L_\mathrm{vac} &= \label{eq:MFlagr-chi} \\
&\hskip -.61cm
    \begin{pmatrix}
        \overline N_1, \overline N_2
    \end{pmatrix} \!
    \begin{pmatrix}
        i\slashed{\partial} - g_1 f_\pi  &  \gamma_5 ( 2 h f_\pi\,  i\slashed{\partial} -m_0 ) \\
       \gamma_5 ( 2 h f_\pi \, i\slashed{\partial} + m_0 )  & i\slashed{\partial} - g_2 f_\pi 
    \end{pmatrix}\!
    \begin{pmatrix}
          N_1\\
          N_2
   \end{pmatrix} . \nonumber
\end{align}
This quadratic form defines the Nambu-Gorkov Dirac operator, $S^{-1}(p) $ in momentum space, of the two-flavor parity-partner baryons $(N_1, N_2)$ for constant $\sigma $ and/or pion fields. Since $\gamma_0 S^{-1}(p) $ is hermitian, it can of course be diagonalized by a unitary transformation which here, for $h\not=0$, necessarily depends on the four-momentum $p$ of the baryon, however, in addition to the meson field expectation value $\esig=f_\pi $ (and the constants $m_0$, $g_1$, $g_2$, $h$). In order to find the propagating normal modes corresponding to the physical mass and parity eigenstates, it turns out to be sufficient to apply a non-unitary but momentum-independent baryon-field transformation of a form, generalizing that in \eqref{eq:PhysFields}, 
\begin{equation}
   \begin{pmatrix}
          N_+\\
          N_-
   \end{pmatrix}
 = F
   \begin{pmatrix}
          N_1\\
          N_2
   \end{pmatrix} , 
\;\mbox{with}\;\; 
   F
 \equiv \begin{pmatrix}
          \cos\theta_1         &  \gamma_5\sin\theta_2 \\
         -\gamma_5\sin\theta_1  &  \cos\theta_2
    \end{pmatrix}
  \, . \label{eq:defFieldTrafo}
\end{equation}
This transformation is not measure preserving, but one has $\mathrm{det}\, F = \cos(\theta_1-\theta_2) $ which introduces a meson-field dependent Jacobian, as we will see below. Moreover, we will need
\begin{equation}
    F^\dagger F = \begin{pmatrix}
        1 & - \gamma_5 \sin(\theta_1-\theta_2) \\ 
       - \gamma_5 \sin(\theta_1-\theta_2) & 1
    \end{pmatrix} \, , \label{eq:FdaggerF}
\end{equation}
and  
\begin{equation}
    \overline F \equiv \gamma_0 F^\dagger \gamma_0 = \left. F^\dagger \right\vert_{\gamma_5 \to - \gamma_5} \,. 
\end{equation}
With these definitions, we express the vacuum Lagrangian in terms of the physical baryon fields as 
\begin{align}
    \mathcal L_\mathrm{vac} =&\, \begin{pmatrix}
        \overline N_+, \overline N_-
    \end{pmatrix} 
    \begin{pmatrix}
        i\slashed{\partial} - m_+ &  0 \nonumber \\
       0  & i\slashed{\partial} - m_- 
    \end{pmatrix}
    \begin{pmatrix}
          N_+\\
          N_-
   \end{pmatrix}\\ 
   =&\,  \begin{pmatrix}
        \overline N_1, \overline N_2
    \end{pmatrix} 
        i\slashed{\partial} \, F^\dagger F \,    \begin{pmatrix}
          N_1\\
          N_2
   \end{pmatrix} \\
&\hskip .8cm   -   \begin{pmatrix}
        \overline N_1, \overline N_2
    \end{pmatrix} 
       \, \overline F \, \begin{pmatrix}
    m_+ &  0 \\
       0  & m_- 
    \end{pmatrix} F \,    \begin{pmatrix}
          N_1\\
          N_2
   \end{pmatrix}
   \, . \nonumber
\end{align}
Comparing this with \eqref{eq:MFlagr-chi}, and using \eqref{eq:FdaggerF}, we can first identify
\begin{equation}
    \sin(\theta_1-\theta_2) = 2 h f_\pi  \, . \label{eq:angles_1}
\end{equation} 
To diagonalize the mass matrix, we furthermore need
\begin{equation}
    \overline F \, \begin{pmatrix}
    m_+ &  0 \\
       0  & m_- 
    \end{pmatrix}  F  \,\stackrel{!}{=} \,
    \begin{pmatrix}
        g_1 f_\pi  &  \gamma_5 m_0  \\
       -\gamma_5 m_0   & g_2 f_\pi  
    \end{pmatrix}\, .
\end{equation}
Decomposing $m_\pm = \overline m \pm \Delta m $ as in the previous section, this leads to
\begin{align}
    g_1f_\pi =&\, \Delta m +\overline m \,\cos 2\theta_1\, , \nonumber\\
    g_2 f_\pi =&\, -\Delta m + \overline m \, \cos 2\theta_2\, , \label{eq:angles_2}\\
        m_0 =&\, \overline m \, \sin(\theta_1+\theta_2) - \Delta m\, \sin(\theta_1-\theta_2)\, . \nonumber
 \end{align}
If the parameters $m_0$, $g_1$, $g_2$, and
$h$ are determined from QCD as the underlying microscopic theory, then one can invert the four relations in \eqref{eq:angles_1} and \eqref{eq:angles_2} to determine the two masses $m_+$, $m_-$
and the two mixing angles
$\theta_1$ and $\theta_2$. In practice, however, one uses phenomenological input, e.g.\ for the masses, and wants to adjust
the ``original'' parameters accordingly. Therefore, we do not intend to invert the relations in
\eqref{eq:angles_1} and \eqref{eq:angles_2}, but instead express
observables in terms of the mixing angles and the physical masses. Below, we will discuss various ways to determine the values of the two mixing angles
from different observables. Before that, we will first provide an alternative equivalent Hamiltonian formulation, which illustrates the role of the normal modes as the baryonic quasiparticle excitations from a different and slightly more general angle.    

\subsection{Nambu-Gorkov Dirac Hamiltonian}
\label{sec:NGDH}

The strategy to find the baryonic normal modes is to first obtain the Nambu-Gorkov-Dirac (single-particle) Hamiltonian corresponding to the Lagrangian, say in Eq.~\eqref{eq:Lagr-ext}, with constant meson-field expectation value. In slightly more generality, we can thereby assume an arbitrary vacuum alignment with a  spacetime constant meson matrix $M$ to represent a general  $O(4)$ vector of $(\sigma,\vec\pi)$ as in Appendix \ref{sec:MFL-VA}, corresponding to the \emph{mean-field} Lagrangian in \eqref{eq:MFlagr-chi}.  Writing $M = \phi U$ with
\begin{equation}
   U = \hat\sigma+i\gamma_5 \hat\pi\, , \;\; \hat\sigma = \sigma/\phi\, , \;\; \hat\pi = \vec\tau\vec\pi/\phi  \, ,\;\; \phi^2=\sigma^2 +\vec\pi^2\, ,  
   \nonumber
\end{equation}
where $U = \mathds 1 $ in the vacuum with standard alignment, we first read off the inverse Nambu-Gorkov propagator $S^{-1}$ from the Lagrangian for constant  $\sigma$ and/or $\vec\pi$ meson fields, which yields in momentum space
\begin{align}
    (\gamma_0 S^{-1})_{11} &= \omega - \gamma_0 \vec\gamma\vec p 
    - g_1 \phi \gamma_0 U \, , \nonumber \\  
    (\gamma_0 S^{-1})_{12} &= -2h \phi \gamma_5 U^\dagger \big(\omega - \gamma_0 \vec\gamma\vec p \big)
    - m_0\gamma_0\gamma_5 \, , \\
    (\gamma_0 S^{-1})_{21} &= -2h\phi \gamma_5 U \big(\omega - \gamma_0 \vec\gamma\vec p \big)
   +  m_0\gamma_0\gamma_5 \, , \nonumber \\ 
    (\gamma_0 S^{-1})_{22} &= \omega - \gamma_0 \vec\gamma\vec p 
    - g_2  \phi \gamma_0 U^\dagger \, 
    .\nonumber
\end{align}
To find the corresponding Nambu-Gorkov Dirac Hamiltonian $H$,  we have to diagonalize the frequency dependent part, and rescale the diagonal blocks so that $\gamma_0 S^{-1} \to  \omega - H $. This is achieved by first applying the unitary transformation 
\begin{equation}
    G = \frac{1}{\sqrt{2}} \begin{pmatrix}
     1 & -\gamma_5 U^\dagger\\
     \gamma_5 U & 1
        \end{pmatrix} \,,\;\; 
         G^\dagger = \frac{1}{\sqrt{2}} \begin{pmatrix}
     1 & \gamma_5 U^\dagger\\
  -  \gamma_5 U & 1
        \end{pmatrix} \,,
\end{equation}
which yields
\begin{widetext}
\begin{align}
    G(\gamma_0 S^{-1}) G^\dagger &=\\[4pt]
    &\hskip -1cm
    \begin{pmatrix}
        (1+2h\phi) (\omega - \gamma_0\vec\gamma\vec p )  - \frac{1}{2} (g_1-g_2) \phi \gamma_0 U + m_0 \gamma_0 U & - \gamma_0\gamma_5 \frac{1}{2} (g_1+g_2 )\phi \\
       \gamma_0\gamma_5 \frac{1}{2} (g_1+g_2 )\phi  & 
         (1-2h\phi) (\omega - \gamma_0\vec\gamma\vec p )  + \frac{1}{2} (g_1-g_2) \phi \gamma_0 U^\dagger  + m_0 \gamma_0 U^\dagger 
    \end{pmatrix} \nonumber \, .
\end{align}
\end{widetext}
The second step is rescaling 
\[ G(\gamma_0 S^{-1}) G^\dagger \to D^{-1/2} G(\gamma_0 S^{-1}) G^\dagger D^{-1/2} \]  
with the block-diagonal matrix 
\begin{equation}
D = \begin{pmatrix}
1+2h\phi & 0 \\
0 & 1-2h\phi 
\end{pmatrix} \, ,
\end{equation}
so that $D^{-1/2} G(\gamma_0 S^{-1}) G^\dagger D^{-1/2} \equiv \omega  - H $, which defines the Nambu-Gorkov Dirac Hamiltonian whose blocks are then given by
\begin{align}
    H_{11} =& \gamma_0 \vec\gamma\vec p +\frac{1}{1+2h\phi}
    \Big( \frac{1}{2} (g_1-g_2) \phi \gamma_0 U -  m_0 \gamma_0 U \Big) \, ,   \nonumber \\
    H_{12} =&  \frac{1}{\sqrt{1-4 h^2\phi^2 }} \, \gamma_0\gamma_5 \frac{1}{2} (g_1+g_2 )\phi    \, , \label{eq:NGD}\\
    H_{21} =& -\frac{1}{\sqrt{1-4 h^2\phi^2 }} \, \gamma_0\gamma_5 \frac{1}{2} (g_1+g_2 )\phi   \, , \nonumber \\
    H_{22} =& \gamma_0 \vec\gamma\vec p 
    \nonumber \\ & {}
    + \frac{1}{1-2h\phi } 
\Big(-\frac{1}{2} (g_1-g_2) \phi \gamma_0 U^\dagger -  m_0 \gamma_0 U^\dagger \Big)  \, .  \nonumber
\end{align}
Because $\gamma_0 U = U^\dagger \gamma_0 $ the Nambu-Gorkov Hamiltonian $H$ is hermitian, with 
\[ H_{11}^\dagger = H_{11} \, ,\;\;  H_{22}^\dagger = H_{22} \, , \;\;\mbox{and} \;\; H_{12}^\dagger = H_{21} \, .\]
The block-diagonal kinetic term $\gamma_0 \vec\gamma\vec p  $ is of the standard form now, and the mass matrix can be brought into block-diagonal form as well, with a momentum-independent unitary transformation  $H\to V H V^\dagger$ which leaves the kinetic term invariant, using
\begin{equation}
    V = \begin{pmatrix}
        \cos\varphi & \gamma_5 \sin\varphi \, U^\dagger \\
        -\gamma_5 \sin\varphi & \cos\varphi  \, U^\dagger
    \end{pmatrix} \, .
\end{equation}
The condition that the off-diagonal blocks vanish afterwards then leads to
\begin{align}
    \big(-\gamma_5 H_{11} + U^\dagger H_{22}\gamma_5 U\big) \cos\varphi\sin\varphi 
    &=  \\
    \gamma_5 H_{12}\gamma_5 U \sin^2\varphi &- U^\dagger H_{21} \cos^2\varphi \, . \nonumber
\end{align}
Working this all out, 
we obtain here
\begin{equation}
    \tan 2\varphi  = - \frac{(g_1+g_2) \phi }{2m_0 + 2h\phi (g_1-g_2) \phi  } \sqrt{1-4h^2\phi^2} \, . \label{eq:varphi}
\end{equation}
For $h=0$, i.e.~without the kinetic mixing, this agrees with Eq.~\eqref{eq:theta_mix} above. To see this, remember that we have implicitly already rotated by an angle $\theta_0 = \pi/4$ in the opposite direction with $G $ in step one which we have to undo to compare with the $h=0$ case above. Thus, using $\varphi = \theta +\pi/4   $ for the rotation remaining in the last step, we can write $\tan 2\varphi  = \tan(2\theta + \pi/2) = - \cot 2\theta  $ and hence
\begin{equation}
    \tan 2\theta =\frac{2m_0 + 2h\phi (g_1-g_2) \phi  }{(g_1+g_2) \phi \,\sqrt{1-4h^2\phi^2} }\, . 
\end{equation}
The diagonal blocks in the Nambu-Gorkov Dirac Hamiltonian are then obtained from
\begin{align}
    H_+ &= H_{11} \cos^2\varphi +U^\dagger \gamma_5H_{22}\gamma_5 U \sin^2\varphi
    \label{eq:Hpgen} \\
  &  \hskip .6cm + \big( H_{12} \gamma_5 U +U^\dagger \gamma_5 H_{21}\big) \cos\varphi\sin\varphi\, ,\nonumber\\
     H_- &= \gamma_5 H_{11} \gamma_5 \sin^2\varphi + U^\dagger H_{22} U \cos^2\varphi \label{eq:Hmgen} \\
     & \hskip .6cm -  \big(\gamma_5  H_{12} U  +U^\dagger H_{21} \gamma_5\big) \cos\varphi\sin\varphi\, .\nonumber
\end{align}
Both blocks are of the form 
\begin{equation}
    H_\pm = \gamma_0 \vec\gamma\vec p + m_\pm \gamma_0 U \, ,
\end{equation}
where the masses are worked out in Appendix~\ref{sec:MFL-VA} as
\begin{align}
    m_+ =& \label{eq:Hamiltonian_massp}\\
    & \frac{1}{2}(g_1-g_2) \phi + \frac{1}{2}(g_1+g_2)\phi \, \frac{1-2h \phi \cos 2\varphi}{\sqrt{1-4h^2\phi^2}\, \sin 2\varphi} \,, \nonumber\\
    m_- =&  \label{eq:Hamiltonian_massm}\\
    &- \frac{1}{2}(g_1-g_2) \phi + \frac{1}{2}(g_1+g_2)\phi\,  \frac{1+ 2h \phi \cos 2\varphi}{\sqrt{1-4h^2\phi^2}\, \sin 2\varphi } \,.
    \nonumber
\end{align}
This in turn implies
\begin{align}
    m_+ - m_- =& \, (g_1-g_2) \phi  \\
    &\hskip .4cm - \frac{2h\phi}{\sqrt{1-4h^2\phi^2}}  \, (g_1+g_2)\phi \cot 2\varphi \nonumber\\
    m_+ + m_- =& \, \frac{(g_1+g_2)\phi }{\sqrt{1-4h^2\phi^2}\, \sin 2\varphi } 
    \, .
\end{align}
We can use these to find the relatively simple relation
\begin{equation}
    m_0 = - \overline m \, \cos 2\varphi - 2h\phi \, \Delta m 
    = \overline m \, \sin 2\theta  - 2h\phi \, \Delta m 
    \, ,
\end{equation}
where $\overline m = (m_++m_-)/2$ and $\Delta m = (m_+-m_-)/2$ as in Sec.~\ref{sec:standard} above. Comparing this relation with Eq.~\eqref{eq:angles_2}, for $\phi = \esig=f_\pi$, we can hence identify
$\varphi-\pi/4 = \theta \equiv (\theta_1+ \theta_2)/2$ and $2 h\phi \equiv \sin(\theta_1-\theta_2 )$, in agreement with Eq.~\eqref{eq:angles_1}. The latter then also implies   
\begin{align}
    g_1 \phi &= \Delta m + \overline m \, \cos 2\theta_1 \, ,\\
     g_2 \phi &= - \Delta m + \overline m\,  \cos 2\theta_2 \, ,
\end{align}
which completes the map of our Hamiltonian formulation, generalized for arbitrary vacuum alignment here, to all four relations in \eqref{eq:angles_1} and \eqref{eq:angles_2}. In particular, this map also implies that $|2h\phi | \le 1$. This restriction is reasonable because the Nambu-Gorkov Dirac operator in presence of the kinetic mixing implicitly contains (field dependent) wavefunction renormalization factors $ (1\pm 2h\phi ) $ which must remain positive in a stable ground state with constant $\phi $.

\section{Baryon-meson couplings}
\label{sec:baryon-meson-couplings}

The diagonalization procedure, applied in the previous section to find the masses of the physical states, can also be used for the determination of the meson-baryon three-point coupling constants. Details can be found in Appendix \ref{sec:details} where we distinguish between the (pseudo-)scalar couplings caused by the $g_1$ and $g_2$ terms in Eq.~\eqref{eq:Lagr-ext} and the derivative couplings caused by the terms $\propto  h$ and $\Delta h$ in \eqref{eq:extLagr}. In Section \ref{sec:results}, we discuss the prospects to determine part of our model parameters by comparison to experimental data on baryon decays (beta decay of the neutron for $g_A$ and strong decays of the resonance $N^*$ for $g_{\pi NN^*}$ and $g_{\sigma N N^*}$). When addressing these decays by a tree-level calculation, all baryons are treated as on-shell states. One can thus use free equations of motion to rewrite derivative couplings (e.g.\ the standard pseudo-vector coupling) into (pseudo-)scalar couplings. 

After this process, we obtain from the results of Appendix \ref{sec:details} the following three-point interaction terms for the pions: 
\begin{align}
  {\cal L}_\pi &=  - \frac{m_+}{f_\pi} \, g_{A}^{++} \, \overline{N}_+ \, i \vec\tau \cdot \vec \pi \gamma_5 N_+ \nonumber\\
  &\hskip .8cm + \frac{m_-}{f_\pi} \, g_{A}^{--} \, \overline{N}_- \, i \vec\tau \cdot \vec \pi \gamma_5 N_-  \label{eq:lagrpionbar-w-eom}   \\
  & {}- \frac{m_--m_+}{2 f_\pi} g_{A}^{+-} \left( \overline{N}_+ \, i \vec\tau \cdot \vec \pi N_- - \overline{N}_- \, i \vec\tau \cdot \vec \pi N_+ \right)
  \, , \nonumber
\end{align}
with the axial charges
\begin{align}
  g_A \equiv g_{A}^{++} &= \frac{\cos(\theta_1+\theta_2)} {\cos(\theta_1-\theta_2)}
  + 4 \Delta h f_\pi \, \frac{\sin\theta_1 \cos\theta_2 }{\cos^2(\theta_1-\theta_2)}   \,,  \nonumber   \\ 
  g_A^*\equiv g_{A}^{--} &= \frac{\cos(\theta_1+\theta_2)} {\cos(\theta_1-\theta_2)}
                   + 4 \Delta h f_\pi \, \frac{\cos\theta_1 \sin\theta_2 }{\cos^2(\theta_1-\theta_2)}   \,, \label{eq:gAetc-main}  \\
  g_{A}^{+-} 
  &= \frac{\sin(\theta_1+\theta_2)}{\cos(\theta_1-\theta_2)} - 2 \Delta h f_\pi \, \frac{\cos(\theta_1+\theta_2) }{\cos^2(\theta_1-\theta_2)} \, ,
  \nonumber
\end{align}
generalizing Eqs.~\eqref{eq:sPDMgA} and \eqref{eq:sPDMgApm} of the standard PDM in Section~\ref{sec:standard} which are recovered upon setting $\theta_1=\theta_2=\theta $ with $\Delta h = 0$. 
As shown in Appendix \ref{sec:details}, these axial charges are the pertinent coupling strengths with which the baryons couple to the weak left-handed currents. Equation \eqref{eq:lagrpionbar-w-eom} demonstrates then that the Goldberger-Treiman relations are indeed satisfied, i.e. 
\begin{eqnarray}
    g_{\pi NN} &=& \frac{m_+}{f_\pi} \, g_{A} \,, \nonumber \\
    g_{\pi N^*N^*} &=& \frac{m_-}{f_\pi} \, g_A^* \,, \nonumber \\
    g_{\pi NN^*} &=& \frac{m_--m_+}{2f_\pi} \, g_{A}^{+-} \,.
    \label{eq:Goldberger-explicit}
\end{eqnarray}

The corresponding interaction Lagrangian for the $\sigma$ meson reads 
\begin{eqnarray}
    {\cal L}_{\sigma} &=& -g_{\sigma NN} \, \overline{N}_+ \, \tilde\sigma N_+ - g_{\sigma N^*N^*} \overline{N}_- \tilde\sigma  N_- \nonumber \\
    && {}+ g_{\sigma NN^*} \left( \overline{N}_+ \, \gamma_5 \tilde\sigma N_- - \overline{N}_- \, \gamma_5\tilde\sigma  N_+ \right) \, ,
    \label{eq:lagrsigmabar-w-eom}
\end{eqnarray}
with the $\sigma$ meson-baryon couplings
\begin{align}
    g_{\sigma NN} &= \frac{m_+}{f_\pi}\frac{\cos^2(\theta_1+\theta_2) + 2\sin^2\theta_1\cos^2\theta_2}{\cos^2(\theta_1-\theta_2)} \nonumber\\
    & \hskip .4cm - \frac{m_-}{f_\pi}\frac{2\sin^2\theta_1\cos^2\theta_2}{\cos^2(\theta_1-\theta_2)} \, ,\nonumber \\
    g_{\sigma N^*N^*} &= \frac{m_-}{f_\pi}\frac{\cos^2(\theta_1+\theta_2) + 2\sin^2\theta_2\cos^2\theta_1}{\cos^2(\theta_1-\theta_2)} \nonumber\\
    &\hskip .4cm - \frac{m_+}{f_\pi}\frac{2\sin^2\theta_2\cos^2\theta_1}{\cos^2(\theta_1-\theta_2)} \, ,  \label{eq:fin-sigma-coupl} \\
    g_{\sigma NN^*} &= \frac{m_+}{f_\pi}\frac{\cos(\theta_1+\theta_2)
    \left(  \sin(\theta_1+\theta_2) - 2\sin(\theta_1-\theta_2) \right)
    }{2\cos^2(\theta_1-\theta_2)} \nonumber \\
    &  +\frac{m_-}{f_\pi}\frac{\cos(\theta_1+\theta_2)
    \left( \sin(\theta_1+\theta_2)  + 2 \sin(\theta_1-\theta_2) \right)
    }{2\cos^2(\theta_1-\theta_2)} \nonumber \\
    & \hskip .4cm  +\frac{\Delta h \, \left( m_+ + m_- \right)}{\cos(\theta_1-\theta_2)} \,.
   \nonumber
\end{align}
And again, using Eqs.~\eqref{eq:mass_ad} and \eqref{eq:GT_aux}, one can verify that these couplings reduce to those in Eqs.~\eqref{eqs:sigBB} when setting $\theta_1=\theta_2=\theta $ and $\Delta h = 0$ in the standard PDM.

\section{Results}
\label{sec:results}

\subsection{PDM baryon masses in the chiral limit}

\begin{figure*}
    \centering
    \includegraphics[width=0.49\linewidth]{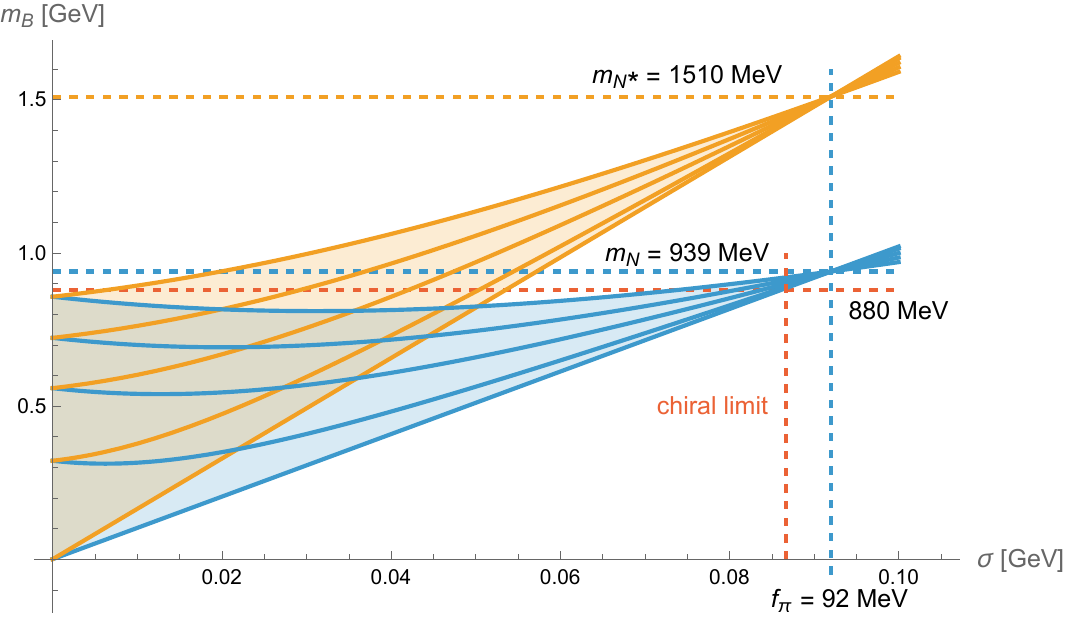}\hfill
    \includegraphics[width=0.49\linewidth]{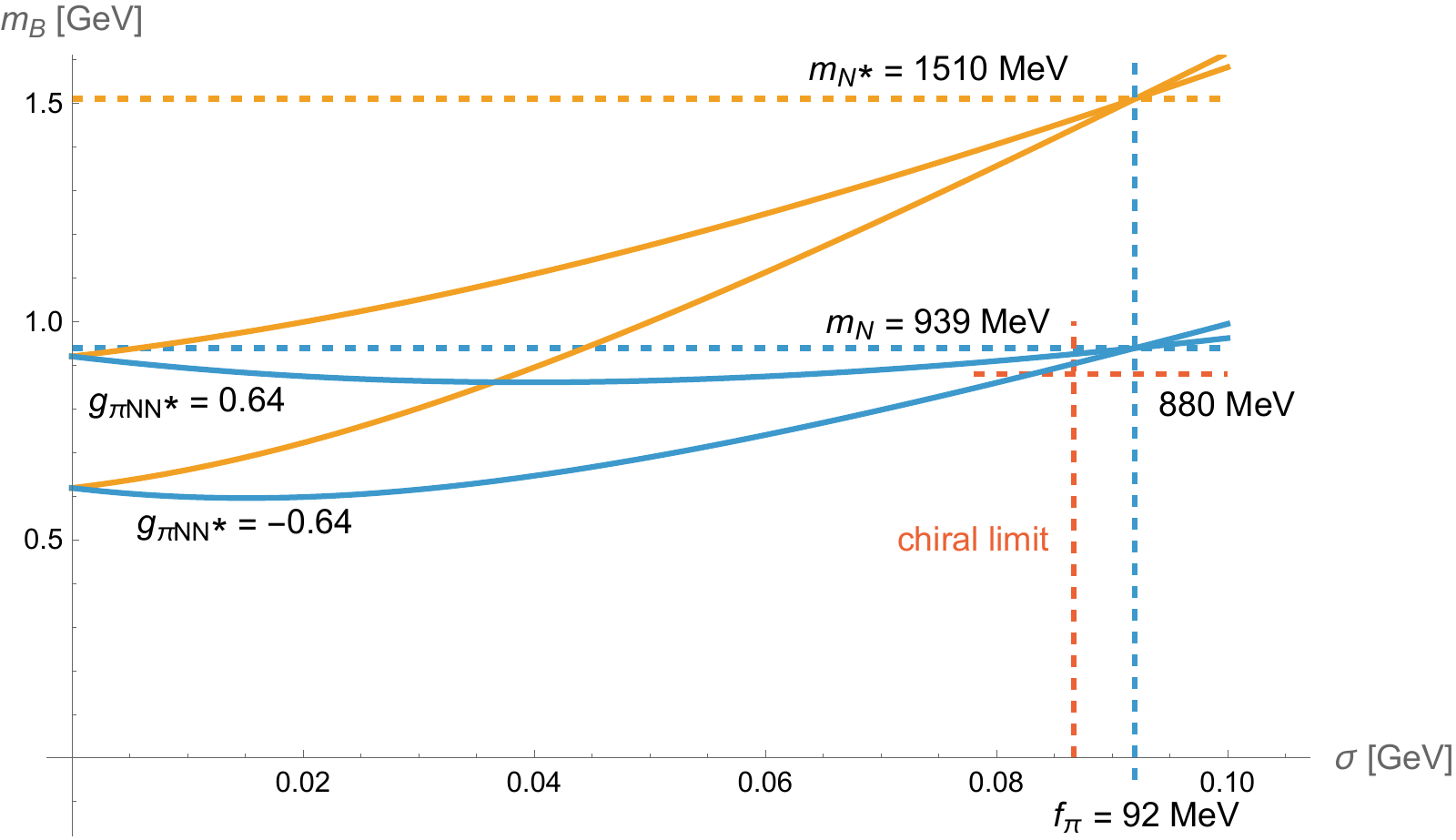}
    \caption{Masses of nucleon (blue) and $N^*(1535)$ (orange) in the standard PDM (with $h=0$) over the $\sigma$-field expectation value. Left: for chirally invariant baryon masses  $m_0 = \{ 0,\, 321, \, 558, \, 722, \, 857\}$~MeV, corresponding to nucleon masses in the chiral limit, where $\langle\sigma\rangle = f_\pi^0 $, of  $m_N^0 = \{ 885,\, 890, \, 900, \, 910, \, 920\}$~MeV (from bottom up). Right: for $m_0 = \{ 618,\, 920 \}$~MeV, such that $|g_{\pi NN^*}| = 0.64 $ is adjusted to the $N^* \to \pi N$ decay as discussed in Sec.~\ref{sec:matchpd-coupl}, leading to the special parameter sets in gray with  $m_N^0 = \{ 903, \, 925\}$~MeV in Tab.~\ref{tab:PDM-Masses-NoKM}.}
    \label{fig:PDM-Masses-NoKM}
\end{figure*}

Explicit chiral-symmetry breaking is introduced in the PDM as in the original Gell-Mann--Levy model by adding an extra symmetry-breaking term $\sim c \sigma $ to the otherwise $O(4)$-invariant mesonic effective potential $V(\phi^2)$. Keeping the parameters in the baryon Lagrangian fixed, the effect of the explicit breaking is to change the pion decay constant from its value in the chiral limit, for $c=0$, to $\langle\sigma\rangle = f_\pi = 92$~MeV at the physical pion mass. In the other direction, the only effect on the baryon masses is then due to the small reduction in the $\sigma$ vacuum expectation value from  $f_\pi$ down to
\begin{equation}
f_{\pi}^{0} = f_\pi/1.061 \approx 87~\mbox{MeV}   
\label{eq:chiral-limit-fpi}
\end{equation}
when going to the chiral limit \cite{FlavourLatticeAveragingGroupFLAG:2021npn}. In the Gell-Mann--Levy model with $m_N = m_1 = g_1 \langle\sigma\rangle  $, and analogously $m_{N^*} = m_2 = g_2 \langle\sigma\rangle$ (for $m_0 = 0$ in the standard PDM), this then immediately implies a corresponding reduction of the nucleon and $N^*$ masses in the chiral limit,
\begin{align}
    m_{N}^0 &= m_N/1.061 \approx  885~\mbox{MeV}\, , \;\;\mbox{and} \nonumber\\
    m_{N^*}^0 &= m_{N^*}/1.061 \approx  1423~\mbox{MeV}\, .
    \label{eq:GMbound}
\end{align}
For the chiral-limit nucleon mass this is at the upper end of current error estimates around $m_{N}^0 \approx 880 $~MeV \cite{Owa:2023tbk,Procura:2003ig,Bernard:2003rp}. At the same time, the Gell-Mann--Levy model estimate in \eqref{eq:GMbound} provides lower bounds on $m_{N}^0$ and $m_{N^*}^0$, in the standard PDM without kinetic mixing. Increasing the chirally invariant baryon mass $m_0 $ inevitably also increases the chiral-limit baryon masses and hence introduces some tension when comparing the standard PDM result to chiral expansions in this way.  
This is demonstrated in Fig.~\ref{fig:PDM-Masses-NoKM}, where we plot the masses of the nucleon and the $N^*(1535)$ in the standard PDM (with $h=0$), for exemplary choices of the chirally invariant baryon mass $m_0$, as functions of the $\sigma$-field expectation value, fixing $g_1 $ and $g_2$ so that  $m_N= 939\,$MeV and $m_{N^*} = 1510\,$MeV at the physical point, where   $\langle\sigma\rangle = f_\pi = 92\,$MeV. The straight lines with $m_0=0 $ correspond to the Gell-Mann--Levy model results and provide the lower bounds for comparison.
The non-vanishing values of $m_0$ are thereby adjusted to yield chiral-limit nucleon masses of $m_N^0 =  890$~MeV,  900~MeV, 910~MeV and  920~MeV, respectively. Since $h=0$, the axial charges of nucleon and $N^*(1535)$ are bound to be identical, $g_A = g_A^* $, in all these results. 
What we can do however, without affecting the baryon-mass mixing of the standard PDM, is to fix the axial charge of the nucleon, for $\theta_1=\theta_2=\theta $ but  $\Delta h \not= 0$,
\begin{equation}
    g_A = g_A^* = \cos 2\theta + 2 \Delta h f_\pi \sin 2\theta \, ,  \label{eq:gA-NoKM}
\end{equation}
by adjusting the derivative coupling $\Delta h$ to obtain $g_A = 1.28$, where the two terms here represent the separate contributions from pseudo-scalar and pseudo-vector pion-baryon couplings, individually. 
Therefore note that in this case, the pseudo-scalar pion-nucleon coupling of the standard PDM remains unchanged in its form,
\begin{equation}
    g_{\pi NN}^s = \frac{m_N}{f_\pi} \, \cos 2\theta \, ,
\end{equation}
and the total pion-nucleon coupling is fixed to $g_{\pi NN} = g_A \, m_N/f_\pi \approx 13.1 $ here simply by adding a pseudo-vector pion-nucleon coupling $f_{\pi NN}$ of suitable strength. Conventionally expressed in units of the pion mass the latter is here identified as 
\begin{equation}
    f_{\pi NN} = \Delta h \, m_\pi \, \sin 2\theta \, . 
\end{equation}
Moreover, the $N^* \leftrightarrow \pi N $ transition coupling $g_{\pi NN^*} $ then follows from $g_A^{+-}$ via the generalized Goldberger-Treiman relation in \eqref{eq:Goldberger-explicit}, which with \eqref{eq:gAetc-main} for $h=0$ reduces to,
\begin{equation}
    g_{\pi NN^*} = \frac{m_{N^*} -m_N}{2f_\pi} \, \big(\sin 2\theta - 2\Delta h f_\pi \cos 2\theta \big) \, . 
\end{equation}
Analogously, the $N^* \leftrightarrow \sigma N $ transition coupling $g_{\sigma NN^*} $ in the last equation of \eqref{eq:fin-sigma-coupl} for $h=0$ here with the auxiliary equations in \eqref{eq:GT_aux} is shown to reduce to 
\begin{equation}
     g_{\sigma NN^*} = \frac{m_{N^*} + m_N}{f_\pi} \, \big( \sin\theta\cos\theta + \Delta h f_\pi \big) \, .
\end{equation}
The values of $\Delta h $, $g_{\pi NN}^s $, $f_{\pi NN} $  and $g_{\pi NN^*}$,  $g_{\sigma NN^*}$ obtained in this subsection without kinetic mass mixing ($h=0$) but with pseudo-vector pion-baryon couplings adjusted to $g_A = 1.28$ are summarized in Table~\ref{tab:PDM-Masses-NoKM}.
The phenomenological estimates of the magnitudes of $g_{\pi NN^*}$ and $g_{\sigma NN^*}$ are discussed next.

\begin{table*}[t]
    \centering
    \begin{tabular}{r|r|r|r||r|r|r||r|r|r|r}
           $m_0$ [GeV] & \hspace{.4cm} $g_1$ & \hspace{.4cm} $g_2 $ &  $\Delta h $ [GeV$^{-1}$] & $m_N^0$ [GeV] & $m_{N^*\,}^0$[GeV] &  \hspace{.4cm} $g_A^0$ & \hspace{.1cm} $g_{\pi NN}^s $ & $f_{\pi NN} $ &  $g_{\pi NN^*}$  &$g_{\sigma NN^*}$  \\
         \hline\hline 
         0 & 10.21 & 16.41 & $\infty $ & 0.885 & 1.423 & 1.28 &  10.21  & 0.212  & $-\infty$ & $\infty $   \\
        0.321 & 9.74 & 15.95 & 6.524 & 0.890 & 1.428 & 1.27 & 9.85 & 0.239 & $-2.78$ & 19.4  \\
         0.558 & 8.74 & 14.95 &  4.649 & 0.900 & 1.438 & 1.26 & 9.08 & 0.296 & $-0.947$ & 16.8  \\
            \textcolor{gray}{0.618} & \textcolor{gray}{8.38} &\textcolor{gray}{14.59} & \textcolor{gray}{4.479}  & \textcolor{gray}{0.903} & \textcolor{gray}{1.442} & \textcolor{gray}{1.26} & \textcolor{gray}{8.81} & \textcolor{gray}{0.316} & \textcolor{gray}{$-0.640 $} & \textcolor{gray}{16.8}  \\
          0.722 & 7.64 & 13.85 &  4.353  & 0.910 & 1.448 & 1.25 & 8.24 & 0.359 & $-0.176$ & 17.0 \\
          0.857 & 6.41  & 12.61 &  4.393 & 0.920 & 1.458 & 1.24 & 7.29 & 0.429 & $0.379$  & 17.4  \\
          \textcolor{gray}{0.920} & \textcolor{gray}{5.68} & \textcolor{gray}{11.89} & \textcolor{gray}{4.485}  & \textcolor{gray}{0.925} & \textcolor{gray}{1.463} & \textcolor{gray}{1.24} & \textcolor{gray}{6.74} & \textcolor{gray}{0.470} & \textcolor{gray}{0.640} & \textcolor{gray}{17.6} 
    \end{tabular}
    \caption{Parameters for the standard PDM without kinetic mass mixing ($h=0$), all for $m_N= 939\,$MeV and $m_{N^*} = 1510\,$MeV, and with derivative coupling $\Delta h$ adjusted to fix the axial charge of the nucleon to $g_A = 1.28$ (left);
    resulting baryon masses and axial charge in the chiral limit (middle); and baryon-meson couplings at the physical point with $f_\pi = 92\,$MeV (right): pseudo-scalar and pseudo-vector pion-nucleon couplings $g^s_{\pi NN}$  and $ f_{\pi NN} $, and resulting transition couplings $g_{\pi NN^*}$ and $g_{\sigma NN^*}$). 
    The results shown in the left panel of Fig.~\ref{fig:PDM-Masses-NoKM} correspond to those with $m_N^0 = \{ 885,\, 890, \, 900, \, 910, \, 920\}$~MeV; the two rows in gray correspond to the special parameter sets used in the right panel of Fig.~\ref{fig:PDM-Masses-NoKM}, where the magnitude of $g_{\pi NN^*}$ is adjusted to the $N^* \to \pi N$ decay as discussed in Sec.~\ref{sec:matchpd-coupl}.
    }
    \label{tab:PDM-Masses-NoKM}
\end{table*}

One final aspect of potential interest regarding the chiral limit is the reduction of the axial charge of the nucleon predicted by this model description.  
In the chiral limit, we use $f_\pi^0  \tan 2\theta^0 = 2 m_0/(g_1+g_2) $ and 
\begin{equation}
   g_A^0 = \cos 2\theta^0 + 2 \Delta h f_\pi^0 \sin 2\theta^0 \, , 
\end{equation}
instead of Eq.~\eqref{eq:gA-NoKM} to demonstrate the small effect of reducing $f_\pi \to f_\pi^0$  in our calculations. This implies that for $m_0 = 0 $ in the Gell-Mann--Levy model limit, $\theta,\, \theta^0  \to 0 $, and $f_\pi^0 \sin 2\theta^0 \to  f_\pi \sin 2\theta  $, so that $g_A^0 \to g_A $ with $2 \Delta h f_\pi \sin 2\theta \to 0.28$ by definition in this limit. With increasing chirally invariant baryon mass $m_0 > 0 $, however, we obtain a small but increasing amount  of quenching of the axial charge of the nucleon towards the chiral limit, as shown in the column after the chiral-limit baryon masses of Table~\ref{tab:PDM-Masses-NoKM}. 

Note that by the same argument, chiral-symmetry restoration, with $\langle\sigma\rangle \to 0$  at any finite $m_0 > 0 $, on the other hand, requires $\theta \to\pi/4 $, and hence necessarily a complete quenching of the axial charges of both baryons, $g_A = g_A^* \to 0$. This complete quenching with chiral-symmetry restoration remains valid also in presence of kinetic mixing, when $h\not= 0$ and $g_A\not= g_A^*$, as can be inferred from the more general equations in \eqref{eq:gAetc-main}. With $2h \langle\sigma\rangle =\sin( \theta_1 -\theta_2) \to 0  $ in the chirally restored phase, these predict 
\begin{equation}
    g_A\, , \; g_A^* \to 0 \, , \; \mbox{but} \; \; g_A^{+-} \to 1 \, .
\end{equation}

\subsection{Matching to meson couplings}
\label{sec:matchpd-coupl}

\begin{figure*}
    \includegraphics[width=0.49\linewidth]{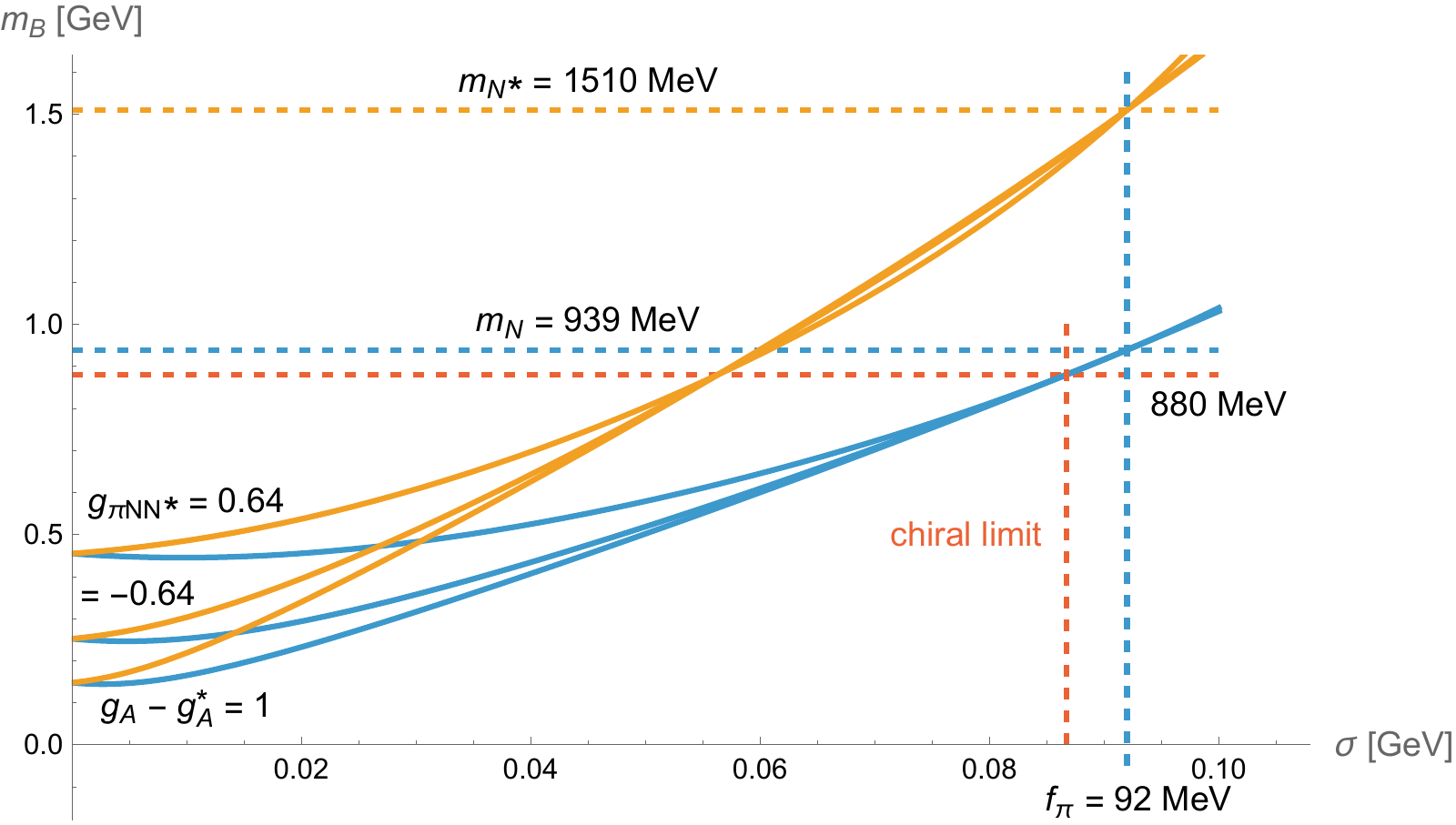}\hfill
    \includegraphics[width=0.49\linewidth]{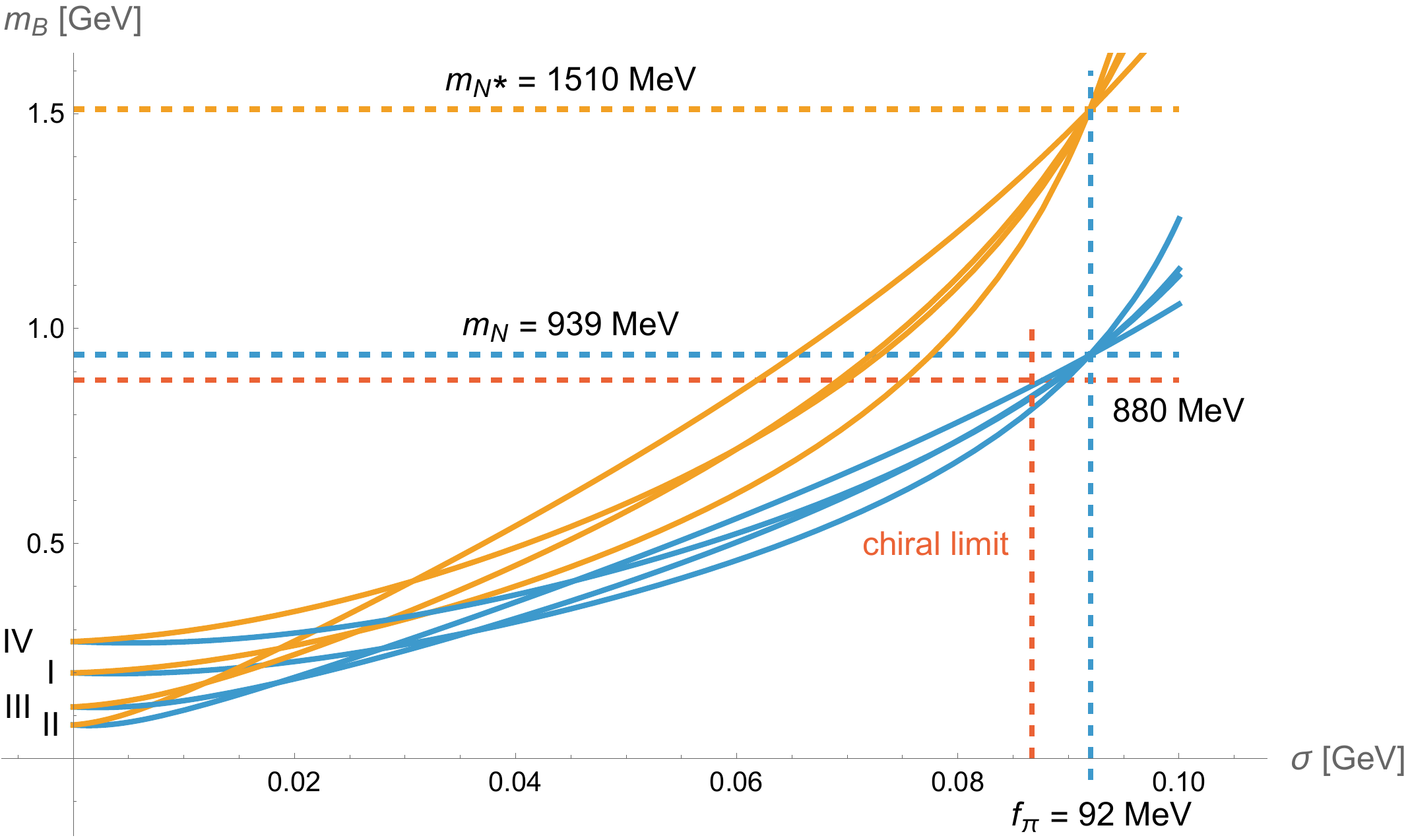}
    \caption{Masses of nucleon (blue) and $N^*(1535)$ (orange)  over the $\sigma$-field expectation value, including kinetic mixing with couplings $h $ and $\Delta h$ chosen to adjust the axial charge of the nucleon to $g_A =1.28$. Left: with nucleon mass in the chiral-limit fixed to $m^0_N=880\,$MeV, comparing solutions where $g_{\pi NN^*} = \pm 0.64 $ and one where the axial charge of the $N^*$ is fixed from the anomaly matching condition $g_A - g_A^* = 1 $. Right: four solutions with $|g_{\pi NN^*}| = 0.64 $ and  $|g_{\sigma NN^*}| = 5.24 $ discussed in Sec.~\ref{sec:matchpd-coupl}.
    }\label{fig:PDM-Masses-gpiNNs}
\end{figure*}

\begin{table*}[t]
    \centering
    \begin{tabular}{r|r|r|r|r||r|r|r|r||r|r|r|r}
           $m_0$ [GeV] & \hspace{.4cm} $g_1$ & \hspace{.4cm} $g_2 $ & $h$ [GeV$^{-1}$]  & $\Delta h $ [GeV$^{-1}$] &  $m_{N^*\,}^0$[GeV] &  \hspace{.2cm} $g_A^0$ & \hspace{.1cm} $(g_A^*)^0 $ & \hspace{.1cm} $(g_A^{+-})^0 $ & \hspace{.2cm} $g_A^* $ & \hspace{.1cm} $ g_A^{+-}  $ &  $g_{\pi NN^*}$  &$g_{\sigma NN^*}$  \\
         \hline\hline 
         0.455 &  5.47 & 15.35 & 3.23 & 0.30 & 1.387  & 1.22 & 1.14 & 0.26 & 1.18 & 0.21 & 0.64 & $-11.12$\\
         0.252 &  8.20 & 15.62 & 2.37 & 1.43 & 1.404 & 1.25 & 1.00 & $ -0.16 $ & 1.00 & $- 0.21$ & $-0.64 $ & $-8.92 $\\
         0.148 &  10.0  & 14.11 & $-2.05$ & $-6.17$ & 1.408 & 1.22 & 0.35 & 1.42 & 0.28 & 1.52 & 4.72 & 15.82
    \end{tabular}
    \caption{Parameters for solutions in the left panel of Fig.~\ref{fig:PDM-Masses-gpiNNs} with kinetic mixing (left: $m_0$, $g_1$ and $g_2$, plus kinetic mixing and derivative couplings $h$ and $\Delta h$), $N^*$ mass and axial charges in the chiral limit (middle: here all with $m_N = 880\, $MeV), and at the physical $f_\pi = 92\,$MeV (right: all with $g_A = 1.28$, top two rows with $|g_{\pi NN^*}| = 0.64 $ and bottom with $g_A - g_A^* = 1$). }
    \label{tab:PDM-Masses-KM}
\end{table*}

As another approach to determine the parameters of our extended PDM, including kinetic mixing with $h\not= 0$, in this subsection we fit coupling constants to the decay data of the $N^*$ resonance, i.e.\ to the partial decay widths $\Gamma_{N^* \to \pi N}$ and $\Gamma_{N^* \to \sigma N}$. At least, for the decay with the $\sigma$ meson in the final state, this has to be taken with a grain of salt. The $\sigma$ meson is so broad that it can hardly be taken as a Breit-Wigner resonance \cite{Pelaez:2015qba,Heuser:2024biq}. The data, based on coupled-channel partial-wave analyses come with large uncertainties. In the following, we use the central values from the Particle Data Group \cite{ParticleDataGroup:2024cfk} for mass and total width of the $N^*(1535)$ ($m_{N^*} \approx 1.510\,$GeV, $\Gamma_{N^*,\rm tot} \approx 0.11\,$GeV). The branching ratios for the two channels of interest are taken from \cite{CBELSATAPS:2015kka}, leading to Br$(N^* \to \pi N) \approx 52$\% and Br$(N^* \to \sigma N) \approx 6$\%. 

The $\pi N$ decay width of an $N^*$ is given by 
\begin{align}
  \Gamma_{N^* \to \pi N} &=  3 \Gamma_{N^{*+} \to \pi^0 p}   
 \label{eq:width-calc}   \\
  &= \frac{3}{8\pi} \frac{p_{\rm cm}}{m_{N^*}^2} \vert g_{\pi NN^*} \vert^2 \left[ (m_{N^*}+m_N)^2- m_\pi^2 \right]  \, ,
  \nonumber
\end{align}
where the momentum 
of the pion and nucleon in the rest frame of the decaying resonance is obtained as
\begin{align}
  \label{eq:pcm-def}
  p_{\rm cm} &= \\
  &\frac1{2 m_{N^*}} \sqrt{\left[ (m_{N^*}+m_N)^2- m_\pi^2 \right] \left[ (m_{N^*}-m_N)^2- m_\pi^2 \right] } \,.\nonumber
\end{align}

The discussion of the more complicated case of the $N^* \to \sigma N$ decay can be found in Appendix~\ref{sec:sigma-broad}. In particular, the partial decay width is provided in \eqref{eq:width-broad-sigma},
\begin{eqnarray}
 \lefteqn{  \Gamma_{N^* \to (\pi\pi)_{\sigma} N} 
  =  \frac1{8 \pi}  \vert g_{\sigma N N^*} \vert^2 \! \int\limits_{4 m_\pi^2}^{(m_{N^*}-m_N)^2} \!\!\! ds \; \frac1{\cal N}\, {\cal A}(s) }
  \nonumber \\ && \phantom{m}\times
  \frac{\left[ (m_{N^*}+m_N)^2- s \right]^{1/2} \!\!}{2 m^3_{N^*}} 
\left[ (m_{N^*}-m_N)^2- s \right]^{3/2}  \!\!\! ,
\nonumber 
\end{eqnarray}
where $\mathcal A (s) $ is the broad spectral function of the $\sigma$-meson with normalization $\mathcal N$. In Appendix~\ref{sec:sigma-broad} we also provide a simple model description of its two-pion resonance contribution, which describes the corresponding pole parameters and the low-energy pion phase shift reasonably well, as shown in Figure~\ref{fig:spec-sigma-impr}.

With these definitions, we obtain the following estimates for the corresponding coupling constants:
\begin{eqnarray}
  \label{eq:det-c-pheno}
  \vert g_{\pi N N^*} \vert \approx 0.64  
\end{eqnarray} 
and 
\vspace{-.2cm}
\begin{eqnarray}
  \vert g_{\sigma N N^*} \vert \approx 5.24  \,.
    \label{eq:detd2}
\end{eqnarray} 
Of course, these are very rough estimates, especially for $g_{\sigma N N^*}$ where the branching ratio comes with a sizable uncertainty of about
$50 \% $ \cite{Sarantsev:2025lik}, and our estimate in Appendix~\eqref{eq:width-broad-sigma} introduces some additional systematic uncertainty. 
The situation is better for $g_{\pi N N^*}$ in \eqref{eq:det-c-pheno} where the branching ratio is known on the $10 \%$ level \cite{CBELSATAPS:2015kka}. 

It might therefore be reasonable to relax \eqref{eq:detd2} for now, and only implement  \eqref{eq:det-c-pheno} in our parameter fixing.
This can in fact be done as a first exercise without kinetic mixing, i.e.\ for $h=0$  as in Fig.~\ref{fig:PDM-Masses-NoKM}, where we see from Table~\ref{tab:PDM-Masses-NoKM} that the family of curves on the left covers both possible signs in $g_{\pi N N^*} = \pm 0.64 $.  We have therefore used these two possibilities to fix the chirally invariant baryon mass $m_0$ as the only adjustable parameter in this case which leads to $m_0 \approx 620$~MeV for one sign and $m_0 \approx 920$~MeV for the other, cf.\ the rows of Tab.~\ref{tab:PDM-Masses-NoKM} highlighted in gray. These two solutions for the baryon masses  $m_\pm(\sigma ) $ are shown separately in the right panel of Fig.~\ref{fig:PDM-Masses-NoKM}. Especially for $m_0 \approx 620$~MeV the chiral-limit nucleon mass of $m_N^0 \approx 900$~MeV is not absurd, but perhaps still acceptable at this level of accuracy.

\begin{table*}
  \centering
  \begin{tabular}{c||r|r|r|r|r||r|r|r|r|r|r}
    \hline
    solution  \vphantom{$\int^A$} & $m_0$ [GeV] & $g_1$ & $g_2$ & $h$ [GeV$^{-1}$] & $\Delta h$ [GeV$^{-1}$] &
     $g_{\sigma NN}$ & $g_{\pi NN}$ & $g_{\sigma N^*N^*}$ & $g_{\pi N^*N^*}$ & $g_{\sigma NN^*}$ & $g_{\pi NN^*}$   \\ 
    \hline \hline
1  &   $0.198$ &   $7.61$ & $5.78$ & $-4.58$ & $1.474$ & $28.10$ & $13.06$ & $ 64.60$ & $46.82$ & $5.24$ & $-0.64$ \\
2   &  $0.078$ &  $8.30$ & $13.44$ & $3.12$ & $0.439$ & $13.99$ & $13.06$ & $26.32$ & $18.73$ & $5.24$ & $-0.64$ \\
3  &  $0.119$ &   $6.51$ & $11.20$ & $4.05$ & $-0.777$ & $19.69$ & $13.06$ & $41.38$ & $28.86$ & $5.24$ & $0.64$ \\  
4  &  $0.272$ &   $8.72$ & $6.55$ & $-4.23$ & $1.021$ & $20.63$ & $13.06$ & $46.01$ & $33.24$ & $5.24$ & $0.64$ \\
  \end{tabular}
  \caption{Values for PDM parameters and three-point coupling constants. The PDM parameter $m_0$, $g_1$, $g_2$, $h$, and $\Delta h$ have
    been chosen such that $\vert g_{\pi NN^*} \vert$ and $\vert g_{\sigma NN^*} \vert$ agree with phenomenology \eqref{eq:det-c-pheno}, \eqref{eq:detd2}.}
  \label{tab:sol-4-new}
\end{table*}

\begin{table*}
  \centering
  \begin{tabular}{c||r|r|r|r|r||r|r|r|r|r|r}
    \hline
    solution \vphantom{$\int^A$} & $m_N^0$ [GeV] & $m_{N^*}^0$ [GeV] & $g_A^0$ & $(g_A^*)^0$ & $(g_A^{+-})^0$ &
     $g_{\sigma NN}^0$ & $g_{\pi NN}^0$ & $g_{\sigma N^*N^*}^0$ & $g_{\pi N^*N^*}^0$ & $g_{\sigma NN^*}^0$ & $g_{\pi NN^*}^0$  \\ 
    \hline \hline
1  & 0.819 & 1.248 & 1.24 & 2.36 & $-0.0641$ &     $21.18$ & $11.64$ & $ 43.26$ & $33.81$ & $5.87$ & $-0.16$  \\
2   & 0.871 & 1.383 & 1.24 & 1.12 & $-0.172$ &    $13.16$ & $12.44$ & $23.96$ & $17.87$ & $4.71$ & $-0.51$  \\
3  & 0.848 & 1.324 & 1.23 & 1.61 & 0.206 &     $16.95$ & $12.00$ & $33.22$ & $24.49$ & $4.18$ & $0.56$ \\  
4 & 0.845 & 1.306 & 1.22 & 1.79 & 0.253 &  $17.00$ & $11.86$ & $34.82$ & $26.89$ & $5.58$ & $0.67$ \\
  \end{tabular}
  \caption{Continuation of Table \ref{tab:sol-4-new}. Values for baryon masses and three-point coupling constants in the chiral limit. }
  \label{tab:sol-4-new-chiral}
\end{table*}

Including the kinetic mixing introduces the coupling $h\not= 0 $ as one further parameter to adjust. Remembering that we always fix the two physical baryon masses and the axial charge of the nucleon $g_A = 1.28 $ we are then left with two additional parameters (which we can think of as $m_0$ and $h$). From the discussions above, we can choose from four candidates of observables to adjust these. We will explore three representative possibilities:

\begin{enumerate}
\setlength{\itemsep}{0pt}
\renewcommand{\labelenumi}{({\bf\alph{enumi}})}
    \item fix the chiral-limit nucleon mass to $m_N^0= 880$~MeV together with the $N^*\to \pi N$ transition coupling to  $g_{\pi N N^*} = \pm 0.64 $.
    \item fix $m_N^0= 880$~MeV as above, but use 
    $g_A^* = 0.28$ for the axial charge of the $N^*(1535)$.
    \item fix both transition couplings,  $|g_{\pi N N^*}| = 0.64 $ and  $|g_{\sigma N N^*}| = 5.24  $, for  $N^*\to \pi N$ and  $N^*\to \sigma N $.
\end{enumerate}

The baryon masses $m_\pm(\sigma ) $ for cases (\textbf{a}) and (\textbf{b}) are shown in the left plot of Figure~\ref{fig:PDM-Masses-gpiNNs},
the corresponding parameters in Table~\ref{tab:PDM-Masses-KM}.

In particular, in (\textbf{a}), with $h\not= 0$,  it is now possible to have  $|g_{\pi N N^*}| = 0.64 $ and $m_N^0= 880$~MeV at the same time, for $m_0 = 455$~MeV  or $m_0 = 252$~MeV, resulting in somewhat smaller chirally-invariant baryon mass values than with $h=0$ above, in either case. 

On the other hand, with these realistic values for the magnitude of the  $N^*\to \pi N$ coupling  $g_{\pi N N^*} $ as in (\textbf{a}), an axial charge of the $N^*(1510)$ results of similar magnitude as $g_A $ (with $g_A^* = 1.18 $ in one case and  $g_A^* = 1.00 $ in the other). Since $h\not= 0$, in presence of the kinetic mixing $g_A$ and $g_A^*$ need no-longer be equal, but the smallness of $g_A^*$ \cite{Takahashi:2008fy,An:2008tz} is difficult to realize together with such a comparatively small $N^*\to \pi N$ transition coupling. 

For comparison, in (\textbf{b}), we therefore sacrifice the smallness of the magnitude of $g_{\pi N N^*} $ on favor of that of $g_A^*$. 
In lack of a more precise  value, here we more or less arbitrarily choose $ g_A^* = 1 - g_A = 0.28 $. This is well within the phenomenological uncertainties  \cite{Takahashi:2008fy,An:2008tz} and has the appealing feature that our extended PDM with kinetic mixing then matches the ABJ triangle anomaly of QCD for $\pi^0 \to 2\gamma $ (as the Gell-Mann--Levy model does, corresponding to $g_A=1$ and $g_A^* = 0$). Note that with kinetic mixing we can always arrange $g_A - g_A^* = 1 $ to match the triangle anomaly by simply implementing the constraint    
\begin{align}
    \Delta h = \frac{1 - 4 h^2 f^2_\pi}{8hf^2_\pi}\,.
    \label{eq:dh-anomaly}
\end{align}
on $h$ and $\Delta h$, from Eqs.~\eqref{eq:gAetc-main}, using $2hf_\pi = \sin(\theta_1- \theta_2)$.

On the other hand, with $g_{\pi N N^*}  = 4.72$ in this case (\textbf{b}), cf.~Tab.~\ref{tab:PDM-Masses-KM}, in presence of kinetic mixing, we observe as a general trend that it is not possible to have a small $g_A^*$ and a sufficiently small  $g_{\pi N N^*} $ (and hence a small $g_A^{+-}$, cf.~Eq.~\eqref{eq:Goldberger-explicit}) at the same time.

Finally, the baryon masses 
 $m_\pm(\sigma ) $ for case (\textbf{c}), where we leave both, the chiral-limit nucleon mass and the axial charge $g_A^*$ of the $N^*(1535)$ resonance unconstrained, are shown in the right  
panel of Figure~\ref{fig:PDM-Masses-gpiNNs}.
There are four independent solutions with  
$|g_{\pi N N^*}| = 0.64 $ and  $|g_{\sigma N N^*}| = 5.24  $. The parameters and the corresponding coupling constants of $N$ and $N^*$ to $\sigma$ and $\pi $ from Eqs.~\eqref{eq:Goldberger-explicit}, \eqref{eq:fin-sigma-coupl} are provided in Table \ref{tab:sol-4-new}.

So far we have glossed over one subtlety when relating the mixing angles, the observables and the PDM parameters. A baryon field, e.g.\ $N$, can be redefined by $N \to -N$ without changing any observable. It just provides an overall quantum mechanical phase that cannot be measured. However, this flips the sign of all coupling constants connecting $N$ to $N^*$. This freedom allows us to arbitrarily fix the sign of one such transition coupling. Without loss, we therefore choose $ g_{\sigma N N^*} > 0$, here. Only the last two columns in Table~\ref{tab:sol-4-new} are influenced by this sign convention. The microscopic PDM parameters would not change if we demanded, e.g., $ g_{\pi N N^*} > 0$. Another way of phrasing this aspect is to note that the mixing angles $\theta_1$ and $\theta_2$ are not uniquely defined by \eqref{eq:angles_1} and \eqref{eq:angles_2}. For instance, a simultaneous change $\theta_1 \to \pi-\theta_1$, $\theta_2 \to 2\pi-\theta_2$ does not change any of these equations. Also those relations for the coupling constants \eqref{eq:Goldberger-explicit}, \eqref{eq:fin-sigma-coupl} will remain the same where only one baryon is involved. However all signs flip for the $N$-$N^*$ couplings. In the mixing formula \eqref{eq:defFieldTrafo}, the upper row of the matrix $F$ receives an extra sign, which is just the operation $N_+ \to -N_+$. 

From the four solutions in Table~\ref{tab:sol-4-new}
solution number 2 stands out as the one with the smallest values for the unconstrained baryon-meson couplings $g_{\sigma NN}$, $g_{\sigma N^*N^*}$ and $g_{\pi N^*N^*}$. This could be important because the matrix elements for the corresponding vertices are proportional to the square of these couplings and might grow too large, too quickly.
In fact, using the Goldberger-Treiman relation for the $N^*(1535)$ in Eqs.~\eqref{eq:Goldberger-explicit} we find that it is only this solution number 2, with $g_A^* = 1.14  $, which at least leads to $g_A^* < g_A $. On the other hand, although this solution has the smallest magnitude of $h$ which might seem preferable to consider kinetic mixing as a perturbation of the standard PDM, it also has the smallest value of the chirally invariant baryon mass $m_0 $, indicating that kinetic mixing in fact dominates over mass mixing in this case.

Using the same model parameters $m_0$, $g_1$, $g_2$, $h$, and $\Delta h$ (and same sign convention) as in Table~\ref{tab:sol-4-new}, we show in Table \ref{tab:sol-4-new-chiral} the results for baryon masses and coupling constants in the chiral limit where Eq.~\eqref{eq:chiral-limit-fpi} applies. 
It is interesting to see that for all solutions, the value for $g_A^0$ remains very close to the physical value of $g_A \approx 1.28$. This is in qualitative agreement with results from lattice QCD; cf.\ the discussion in \cite{Alvarado:2021ibw}.

\section{Summary and Outlook}
\label{sec:summary}

Among the appealing aspects of the parity-doublet model (PDM), as describing chiral symmetry breaking in presence of a purely gluonic, chirally invariant baryon mass from the scale anomaly in QCD, the smallness of the axial charge $g_A \ll 1$ of the nucleon $N$ stands out as one very unsatisfying feature \cite{Jido:2001nt}. Another potential problem of the standard PDM is the rigid relation $g_A = g_A^*$ between $g_A$ and the axial charge $g_A^*$ of the negative parity partner baryon,  conventionally assumed to be the $N^*(1535)$ resonance. In this paper we have shown that both problems are solved by including derivative couplings of $\sigma $-meson and pions to the parity-partner baryons $N$ and $N^*$. In particular, we have extended the Lagrangian of the standard PDM by all such couplings with one derivative that are allowed by symmetry, most importantly by chiral symmetry. This introduces two new couplings of mass dimension $[-1]$, $h_{1,2} \equiv h \pm \Delta h$. While the difference $\Delta h$ of the two has no effect on the mixing of the baryonic mass eigenstates $N$ and $N^*$, it can be used to solve the first problem, the smallness of $g_A $, by introducing (pseudo-)vector meson-baryon couplings. To independently adjust $g_A^*$, however, one needs both independent derivative couplings of the extended chiral model, $h$ and $\Delta h $. In particular, $h\not=0 $ then also affects the mass mixing between $N$ and $N^*$.

We have discussed various possibilities to adjust the model parameters entering the effective baryonic Lagrangian with derivative couplings. For any given pair of values of the poorly constrained chiral baryon mass $m_0 $ and $h$, the two standard Yukawa couplings $g_1$ and $g_2 $ are uniquely determined by the masses of $N$ and $N^*$.  In addition, $\Delta h $ can always be adjusted to reproduce the phenomenological value of the nucleon's axial charge, $g_A \approxeq 1.28$, in all parameter sets that we have explored. This is a significant improvement especially when the parity-doublet model is used to calculate neutrino emissivities of dense baryonic matter \cite{Brodie:2025nww}.   

The remaining two model parameters to fix are then $m_0$ and $h$. We have discussed various possibilities to use two out of a set of four possible observables: the chiral-limit nucleon mass $m_N^0$, the axial charge $g_A^*$ of the $N^*(1535)$, and the two meson-baryon transition couplings, $g_{\pi NN^*}$ and  $g_{\sigma NN^*}$, the magnitudes of which we estimate from the corresponding decay widths.
Calculating the respective other two observables as model predictions can then serve to assess the consistency of the effective description. Although the procedure is limited by the considerable uncertainties in several of these observables, we can summarize a few general trends: a sizable value of the chirally invariant baryon mass $m_0$ is difficult to reconcile with the expected reduction of the nucleon mass in the chiral limit. Moreover, the general mixing pattern makes it impossible to have a small axial charge $g_A^*\propto  g_{\pi N^*N^*}$ of the $N^*(1535)$ \cite{Takahashi:2008fy,An:2008tz} and reasonably small meson-baryon transition couplings $g_{\pi NN^*}$, $g_{\sigma NN^*}$  at the same time. While $g_{\sigma NN^*}$   is plagued by considerable uncertainty,  $g_{\pi NN^*} \propto g_A^{+-} $ (from generalized Goldberger-Treiman relations) is actually rather well constrained. In fact, our estimate of $|g_{\pi NN^*}| \approx 0.64 $ leads to a prediction of $|g_A^{+-}| \approx 0.21 $ for the size of the off-diagonal axial charge connecting $N(939)$ and $N^*(1535)$. 

A general feature of the PDM is the complete quenching of the diagonal axial charges of $N$ and $N^*$ in a putative hadronic high-density phase with chiral symmetry restoration, where $g_A,\, g_A^* \to 0 $ but $g_A^{+-}\to 1 $. This general feature remains true in the extended PDM with kinetic mixing introduced here.

The full chiral structure of hadrons --- and not only the (approximate) vector flavor symmetry --- is probed by the weak interaction with its preference for left-handed particles and right-handed antiparticles. Of course, our present developments, which lead to the correct strength of $g_A$, allow for a quantitatively reasonable description of the classic weak process, the beta decay of the neutron. In this spirit, it would be interesting to extend the framework from two to three light flavors. In the meson sector, chiral perturbation theory is a standard tool for the understanding of kaon and eta decays \cite{Cirigliano:2011ny,Colangelo:2018jxw,Fang:2021wes}. But in the sector of hyperons, there are severe challenges for the application of chiral perturbation theory to non-leptonic and radiative decays \cite{Bijnens:1985kj,Jenkins:1991bt,Jenkins:1992ab,Neufeld:1992np}. Logarithmic corrections to leading-order results seem to work reasonably well in some channels but fail in others. On the other hand, full next-to-leading-order calculations have very limited predictive power, because they involve too many unknown low-energy constants.

In such a situation, a better microscopic understanding of the structure of baryons might help for a reliable estimate of the respective sizes of the low-energy constants. A three-flavor version of the mirror assignment can provide such a microscopic picture. It is also appealing that the mirror assignment involves not only the ground-state baryons but also resonant states. The weak interaction induces a further mixing of baryon states that have {\em different} numbers of strange quarks. In general, this mixing is the larger, the smaller the energy difference of the strong-interaction eigenstates is. For instance, the mass of the $\Sigma$ baryon lies exactly in the middle between the nucleon and the Roper resonance, and also the $N(1535)$ is only further 100 MeV away. It has often been speculated that these resonances play an import role in weak non-leptonic and radiative decays \cite{LeYaouanc:1978ef,Jenkins:1992ab,Borasoy:1999md,Borasoy:1999nt}. Obviously, the mirror assignment would correlate the weak mixing parameters of a hyperon to ground-state baryons and to resonances.

Just recently, LHCb announced the observation of the most rare baryon decay seen so far, namely the process $\Sigma^+ \to p \, \mu^+ \mu^-$ \cite{LHCb:2025evf}. Unfortunately, a severe high-precision test of the standard model is hampered by the limited predictive power of chiral perturbation theory \cite{Jenkins:1992ab,Neufeld:1992np} and lattice QCD \cite{Erben:2025jkq} for the weak radiative decays. More microscopic input based on a better understanding of the chiral structure of baryons can improve the situation. However, this requires to spell out the concrete chiral representations that enter in the description of three-flavor multiplets of ground-state baryons and low-lying resonances. For instance, a two-flavor chiral structure of a left-handed doublet and right-handed singlet, can be generalized to a left-handed triplet and right-handed anti-triplet or to a left-handed octet and right-handed singlet. Various interpolating currents for baryons with well-defined chiral transformation properties have been formulated in \cite{Chen:2008qv}. It would be interesting to explore the different predictions related to specific choices of chiral representations; see also the discussion in \cite{Gao:2024mew}. By having the classical beta decay under control, our present work can be seen as a starting point to explore the consequences of the mirror assignment for the vacuum phenomenology of weak baryon decays. 

At the same time, weak processes play also an important role for nuclear physics. Here the neutrino emissivity for the cooling of neutron stars might serve as a prominent example \cite{Brodie:2025nww}. With the realistic axial charges of our extended PDM, it will therefore be important, in a first step, to reconsider the phenomenologically successful description of the thermodynamics and equation of state of symmetric and asymmetric nuclear and $\beta$-equilibrated neutron-star matter  \cite{Eser:2023oii,Eser:2024xil,Recchi:2025pyy}.

Of course, the extension of the formalism to three light flavors allows to study the importance of hyperons for neutron-star physics \cite{Fraga:2023wtd,Gao:2024mew}. An interesting question is here the nature of the in-medium chiral phase transition. Does the restoration take place for three or just for two flavors? As long as a symmetry remains spontaneously broken, it might seem advantageous to use a non-linear realization for the Goldstone bosons \cite{Coleman:1969sm,Callan:1969sn,Scherer:2012xha}. In this case, it is not necessary to spell out the transformation properties of the baryons with respect to the full chiral group. But one must use a chiral vielbein with its derivative couplings to ensure the Goldstone theorem. A non-linear realization is of advantage if one wants to keep the framework as model independent as possible. One stays as ignorant as possible about the chiral properties of baryons. On the other hand, as already discussed, a linear representation is of advantage, if one wants to develop and test a more microscopic picture of baryons. Coming back to the chiral phase transition, it is certainly easier to use a linear representation for the (part of the) symmetry group that gets restored. However, we note in passing that also a non-linear sigma model can have a chiral phase transition \cite{Bochkarev:1995gi}.

\vspace*{-12pt}

\acknowledgments
We gratefully acknowledge discussions with Mattia Recchi, especially on the mean-field thermodynamics of the extended PDM, and with  Rob Pisarski, Fabian Rennecke, J\"urgen Schaffner-Bielich, and Jochen Wambach.  
This research is supported by the Deutsche Forschungsgemeinschaft, Project No.~315477589, the Collaborative Research Center TRR 211, ``Strong-interaction matter under
extreme conditions,'' and by the European Union’s HORIZON MSCA Doctoral Networks programme, under Grant Agreement No.\ 101072344,  AQTIVATE (Advanced computing, QuanTum algorIthms and data-driVen Approaches for science, Technology and Engineering). C.K.~also acknowledges support by the Erasmus+ Traineeship Programme of the European Union.

\appendix

\section{Parity flip by spontaneous symmetry breaking}
\label{sec:flip}

Here we discuss the interesting (but maybe phenome\-nologically irrelevant) finding that spontaneous symmetry breaking can flip the relative parity between the two baryon states around. Since this is a subtle issue we discuss this case in some detail.

Consider a pair of left-handed and a pair of right-handed baryon fields where the members of each pair transform oppositely under (chiral) isospin transformations, namely
\begin{eqnarray}
  B_{1L} & \to & U_L \, B_{1L} \,, \nonumber \\
  B_{1R} & \to & U_R \, B_{1R} \,, \nonumber \\
  B_{2L} & \to & U_R \, B_{2L} \,, \nonumber \\
  B_{2R} & \to & U_L \, B_{2R} \,. 
  \label{eq:chiraltrafos}
\end{eqnarray}
Here the label $L$ ($R$) for the baryon fields refers to left- (right-)handed fermions concerning the Lorentz group. Note the mirror assignment \cite{Detar:1988kn,Jido:2001nt} for $B_2$.  

In a first step, we restrict ourselves to fermions. The most general $SU(2)_L \times SU(2)_R$ invariant Lagrangian with dimension-3 and dimension-4 operators is
\begin{eqnarray}
  {\cal L}_0 &=&  \overline{B_{1L}} i \slashed{\partial} B_{1L} +  \overline{B_{2L}} i \slashed{\partial} B_{2L}
  +  \overline{B_{1R}} i \slashed{\partial} B_{1R} +  \overline{B_{2R}} i \slashed{\partial} B_{2R} \nonumber \\
  && {} - m_1 \overline{B_{1L}} B_{2R} - m_2 \overline{B_{2L}} B_{1R} 
  \nonumber \\ 
  && {} - m_3 \overline{B_{1R}} B_{2L} - m_4 \overline{B_{2R}} B_{1L} \,.
  \label{eq:lagrwomeson}  
\end{eqnarray}
Hermiticity demands $m_1 = m_4^*$ and $m_2 = m_3^*$. Parity symmetry leads to $m_1=m_3$ and $m_2=m_4$. 
This yields
\begin{eqnarray}
  {\cal L}_0 &=&  \overline{B_{1L}} i \slashed{\partial} B_{1L} +  \overline{B_{2L}} i \slashed{\partial} B_{2L}
  +  \overline{B_{1R}} i \slashed{\partial} B_{1R} +  \overline{B_{2R}} i \slashed{\partial} B_{2R} \nonumber \\
  && {} - m_1 \overline{B_{1L}} B_{2R} - m_1^* \overline{B_{2L}} B_{1R}   \nonumber \\ 
  && {} - m_1 \overline{B_{1R}} B_{2L} - m_1^* \overline{B_{2R}} B_{1L} \,.
  \label{eq:lagrwomeson2}  
\end{eqnarray}
If we write the mass parameter $m_1$ as $\vert m_1 \vert e^{i\alpha}$, then we can move the phase into a redefinition of the $B_1$ fields: 
\begin{eqnarray}
  \label{eq:phase-redef}
  e^{-i\alpha} B_{1L} \to B_{1L} \qquad \mbox{and} \qquad e^{-i\alpha} B_{1R} \to B_{1R}  \,.   
\end{eqnarray}
We end up with the most general Lagrangian
\begin{eqnarray}
  {\cal L}_0 &=&  \overline{B_{1L}} i \slashed{\partial} B_{1L} +  \overline{B_{2L}} i \slashed{\partial} B_{2L}
  +  \overline{B_{1R}} i \slashed{\partial} B_{1R} +  \overline{B_{2R}} i \slashed{\partial} B_{2R} \nonumber \\
  && {} - m_0 \overline{B_{1L}} B_{2R} - m_0 \overline{B_{2L}} B_{1R}   \nonumber \\ 
  && {} - m_0 \overline{B_{1R}} B_{2L} - m_0 \overline{B_{2R}} B_{1L} 
  \label{eq:lagrwomeson3}  
\end{eqnarray}
and a positive mass parameter $m_0 \equiv \vert m_1 \vert$.

We stress that $B_1$ and $B_2$ are not the physical fields, i.e.\ they do not correspond to single-particle excitations.
The physical fields are superpositions of $B_1$ and $B_2$. We also stress that we do not know yet about the relative parity between the
physical fields. This can be determined most easily via interactions with mesons. Finally we have not introduced yet the additional mass generation
via spontaneous symmetry breaking.

To this end, we introduce a meson field $M$ that transforms as 
\begin{eqnarray}
  \label{eq:mesontrafo}
  M \to U_L \, M \, U_R^\dagger \,.
\end{eqnarray}
Spontaneous symmetry breaking will be achieved by the replacement $M^a_b \to f_\pi \delta^a_b$. The parity transformation changes $M$ to
$M^\dagger$. In terms of sigma and pion fields one might write $M = \sigma + i \vec \pi \cdot \vec \tau$ with Pauli isospin matrices $\vec \tau$.

The most general chirally invariant Lagrangian with dimension-4 operators containing our baryon fields and the mesons is 
\begin{eqnarray}
  {\cal L}_M &=& -g_1 \overline{B_{1L}} M B_{1R} - g_1^* \overline{B_{1R}} M^\dagger B_{1L}
    \nonumber \\ 
  && {} + g_2 \overline{B_{2L}} M^\dagger B_{2R} + g_2^* \overline{B_{2R}} M B_{2L} \,.
  \label{eq:lagrwithmesons}  
\end{eqnarray}
Parity symmetry demands $g_i^* = g_i$ leading to  
\begin{eqnarray}
  {\cal L}_M &=& -g_1 \overline{B_{1L}} M B_{1R} - g_1 \overline{B_{1R}} M^\dagger B_{1L}
    \nonumber \\ 
  && {} + g_2 \overline{B_{2L}} M^\dagger B_{2R} + g_2 \overline{B_{2R}} M B_{2L} 
  \label{eq:lagrwithmesons2}  
\end{eqnarray}
with two real-valued parameters $g_1$ and $g_2$. It is important to note that the field redefinitions \eqref{eq:phase-redef} do not change
\eqref{eq:lagrwithmesons2}. Thus ${\cal L}_0 + {\cal L}_M$ is the most general Lagrangian compatible with the chiral symmetry $SU(2)_L \times SU(2)_R$ and parity, provided that one ignores higher-dimensional operators.  

We discuss now the two cases of Wigner-Weyl phase and Nambu-Goldstone phase for the chiral symmetry group $SU(2)_L \times SU(2)_R$.
In the former case, the Lagrangian \eqref{eq:lagrwithmesons2} provides only interactions. We need to diagonalize the
Lagrangian \eqref{eq:lagrwomeson3} such that one obtains the standard form of a free Lagrangian for two baryons.
In a second step we will check what this diagonalization implies for the interaction terms of \eqref{eq:lagrwithmesons2}. This will allow us to
determine the parity of the obtained single-particle baryon states. 

As it stands we can merge the left- and right-handed components together and obtain the compact version
\begin{eqnarray}
  {\cal L}_0 &=&  \overline{B}_1 i \slashed{\partial} B_{1} +  \overline{B}_{2} i \slashed{\partial} B_{2}
  - m_0 \overline{B_{1}} B_{2} - m_0 \overline{B_{2}} B_{1} \,.  \phantom{m}
  \label{eq:lagrwomeson5}  
\end{eqnarray}
We introduce new fields via 
\begin{eqnarray}
  B_1 = \frac{1}{\sqrt{2}} \left( B_A + B_B \right) \,, \quad B_2 = \frac{1}{\sqrt{2}} \left( B_A - B_B \right) 
  \phantom{m}  \label{eq:diag1}  
\end{eqnarray}
and find
\begin{equation}
  {\cal L}_0 =  \overline{B_A} i \slashed{\partial} B_{A} +  \overline{B_{B}} i \slashed{\partial} B_{B}
  - m_0 \overline{B_{A}} B_{A} + m_0 \overline{B_{B}} B_{B} \,.
  \label{eq:lagrwomeson6}  
\end{equation}
This is not the standard form of a free Lagrangian yet, because the mass term for $B_B$ has the wrong sign. In this context we recall that the parameter $m$ is positive by construction. Therefore we introduce
\begin{eqnarray}
  \label{eq:diag2}
  B_+ \equiv B_A \,, \qquad B_- \equiv -\gamma_5 B_B
\end{eqnarray}
or in components
\begin{eqnarray}
  \label{eq:diag2a}
  B_{+L} \equiv B_{AL} \,, &\quad & B_{+R} \equiv B_{AR} \,, \nonumber \\  B_{-L} \equiv B_{BL} \,, & \quad & B_{-R} \equiv -B_{BR} \,.
\end{eqnarray}
As desired we obtain the standard form of a free Lagrangian: 
\begin{equation}
  {\cal L}_0 =  \overline{B_+} i \slashed{\partial} B_{+} +  \overline{B_{-}} i \slashed{\partial} B_{-}
  - m_0 \overline{B_{+}} B_{+} - m_0 \overline{B_{-}} B_{-} \,.  \phantom{m}
  \label{eq:lagrwomeson7}  
\end{equation}
The two baryons have the same mass. 
Finally we rewrite the interaction Lagrangian \eqref{eq:lagrwithmesons2} in terms of the physical fields $B_+$ and $B_-$. A very practical way
to do this is the introduction of the hermitian and anti-hermitian part of the meson matrix $M = M_h + M_a$. The hermitian part is invariant
under parity, i.e.\ describes scalar mesons. The anti-hermitian part is odd under parity and describes pseudo-scalars. We get 
\begin{eqnarray}
  && {\cal L}_M = -g_1 \overline{B_{1}} M_h B_{1} - g_1 \overline{B_{1}} M_a \gamma_5 B_{1}
  \nonumber \\ && \phantom{mmm}
                 + g_2 \overline{B_{2}} M_h B_{2} - g_2 \overline{B_{2}} M_a \gamma_5 B_{2}   \nonumber \\
             && = -\frac12 \left(g_1 - g_2 \right) \overline{B_+} M_h B_+ +  \frac12 \left(g_1 - g_2 \right) \overline{B_-} M_h B_-    \nonumber \\
             &&{} + \frac12 \left(g_1 + g_2 \right) \overline{B_+} M_h \gamma_5 B_-
                -  \frac12 \left(g_1 + g_2 \right) \overline{B_-} M_h \gamma_5 B_+     \nonumber \\
             &&{} -\frac12 \left(g_1 + g_2 \right) \overline{B_+} M_a \gamma_5 B_+
                +  \frac12 \left(g_1 + g_2 \right) \overline{B_-} M_a \gamma_5 B_-    \nonumber \\
             &&{} + \frac12 \left(g_1 - g_2 \right) \overline{B_+} M_a B_-
                -  \frac12 \left(g_1 - g_2 \right) \overline{B_-} M_a  B_+  \,.
                \nonumber \\ &&
  \label{eq:lagrwithmesons3}  
\end{eqnarray}
It is obvious that $B_+$ and $B_-$ have opposite parity. This should also be clear from the point of view of chiral symmetry in the Wigner-Weyl
phase. Massive states have chiral partners with opposite parity (except if the massive state is a chiral singlet). If a conserved axial charge acts on a state, the parity flips but the
energy/mass does not change. The interactions have also the same strength for these chiral partners, as one can easily read off
from \eqref{eq:lagrwithmesons3}. 

After this detailed analysis, one can shortcut the discussion if one is only interested in the masses and relative parity of the baryonic
single-particle states: One just determines the eigenvalues of the mass matrix. If both eigenvalues are positive, then the baryons have
the same parity. If the eigenvalues have different signs, then the physical states have opposite parity relative to each other. If both eigenvalues
are negative, one can redefine the baryon fields such that the masses become positive without any change of the physics. The physical states will have the same parity. 

Now we switch to the Nambu-Goldstone phase of spontaneous symmetry breaking. We want to show that there are parameter choices for
$m_0$, $g_1$, and $g_2$ where the two physical baryon states have the {\em same} parity. Replacing $M^a_b \to f_\pi \delta^a_b$ and collecting all
terms quadratic in the fields yields the free Lagrangian 
\begin{eqnarray}
  {\cal L}_{\rm free,SSB} &=&  \overline{B_1} i \slashed{\partial} B_{1} +  \overline{B_{2}} i \slashed{\partial} B_{2}
                              - m_0 \overline{B_{1}} B_{2} - m_0 \overline{B_{2}} B_{1} \nonumber \\
  && {} - g_1 f_\pi \overline{B_{1}} B_{1} + g_2 f_\pi \overline{B_{2}} B_{2} \,.
  \label{eq:freeSSB}
\end{eqnarray}
Therefore we look for the eigenvalues of the mass matrix
\begin{eqnarray}
  \label{eq:massmatrix}
  \left(
  \begin{array}{cc}
    g_1 f_\pi & m_0 \\ m_0 & -g_2 f_\pi
  \end{array}
      \right)   \,,
\end{eqnarray}
which are given by 
\begin{equation}
  \label{eq:mass-final}
  m_{A,B} = -\frac12 f_\pi \left( g_2 - g_1 \right) \pm \sqrt{m_0^2 + \frac14 f_\pi^2 \left( g_1 + g_2 \right)^2}  \,.
\end{equation}

If both $g_i f_\pi$ are relatively small, i.e.\ if the dominant effect comes from the $m_0^2$ term, then we will find one positive and one negative
mass eigenvalue. In this case one introduces new fields \eqref{eq:diag2} and obtains two states with opposite parity. Their masses are given by
\begin{equation}
  \label{eq:mass-final2}
  m_{+,-} = \sqrt{m_0^2 + \frac14 f_\pi^2 \left( g_1 + g_2 \right)^2} \mp \frac12 f_\pi \left( g_2 - g_1 \right)   \,.
\end{equation}
This is the standard scenario of the PDM in the Nambu-Goldstone phase, two non-degenerate baryons with opposite parity and masses given by \eqref{eq:mass-final2}. In the main text we focus on this scenario and extend it by the introduction of kinetic mixing.

However, if the two coupling constants have different sign, e.g.\ $g_1 > 0$ and $g_2 <0$, and if 
\begin{eqnarray}
  \label{eq:ineq}
  g_1 \, (-g_2) \, f_\pi^2 > m_0^2 \,,
\end{eqnarray}
then the two masses given by \eqref{eq:mass-final} will have the same sign (positive for our sign choice of the coupling constants). There is no reason to redefine one of the baryon fields via \eqref{eq:diag2}. The baryons states labeled by $A$, $B$ qualify as single-particle states. For this non-standard scenario, one obtains two non-degenerate baryons with the same parity and masses given by \eqref{eq:mass-final}. 

What happens if we approach the Wigner-Weyl phase of restored chiral symmetry in this non-standard scenario? In other words, what happens if $f_\pi$ becomes smaller? At some point, the condition \eqref{eq:ineq} is no longer satisfied. At the break-even point of $m_0^2 = -g_2 g_1 f_\pi^2$, the smaller of the two masses vanishes. Note that this happens {\em before} chiral symmetry is fully restored with $f_\pi =0$. After this break-even point, the two physical baryon states have opposite parity. The state that was massless at the break-even point has switched its parity relative to the heavier partner state. With altered parity it regains mass until it becomes degenerate with its parity partner at the point of chiral restoration.
It is certainly a funny possibility
that a particle has a vanishing mass {\em before} chiral restoration is reached and has regained a mass at chiral restoration
(and a parity opposite to the vacuum case).

We stress again that we regard it as less likely that this scenario is realized in nature. Nonetheless, it might be interesting to plug in some numbers. If we search for two baryons close in mass and with same parity, the nucleon and Roper ($N(1440)$) come to mind, in particular given the fact that it is not so clear why the first excitation of the nucleon does not have opposite parity relative to the ground state \cite{Lang:2016hnn,Severt:2022jtg}. 

If both states have the same parity (in the vacuum), then we read off from \eqref{eq:mass-final} the mass difference between the two
states. This difference is given by $2 \sqrt{m_0^2 + \frac14 f_\pi^2 \left( g_1 + g_2 \right)^2}$, which places an upper limit
on $m_0$. If one chooses $g_1=-g_2$ and the masses of nucleon and Roper, then one obtains the rather low value of $m_0 \approx 0.25\,$GeV together with $g_1=-g_2\approx 13$.
When the pion decay constant $f_\pi$, i.e.\ the sigma expectation value, has dropped from about 92 to about 20 MeV, then the nucleon mass vanishes
for this choice of parameter values.
Note that this is just a particular model case. There is no reason to choose $g_1=-g_2$. The purpose is just to show a typical value of $m_0$. For the case $g_1 \neq -g_2$, the value of $m_0$ would further shrink below $0.25\,$GeV, if we keep nucleon and Roper as the states for which this non-standard scenario is supposed to hold. In general, the limits of the parameters for this scenario are $m_0 \le (m_A-m_B)/2$ and $m_B \le |g_2| f_\pi \le (m_A+m_B)/2 \le g_1 f_\pi \le m_A$ (if one assumes $0 < |g_2| \le g_1$). The limiting cases are $m_0 = (m_A-m_B)/2$ together with $g_1 f_\pi=-g_2 f_\pi=(m_A+m_B)/2$ and $m_0=0$ together with $g_1 f_\pi = m_A$, $|g_2| f_\pi = m_B$.

\section{Detailed derivation of the extended PDM Lagrangian}
\label{sec:details}
In this appendix, we formulate a PDM version that contains kinetic mixing. We focus on the baryon sector, i.e.\ we do not spell out explicitly the meson Lagrangian of the linear sigma model that leads to spontaneous symmetry breaking. In the baryon sector, we want to include {\em all} possible terms with no or one derivative and with no or one meson field. We include in the general construction also the coupling to weak currents with the purpose to determine the axial charge $g_A$. We also regard it as useful to check the Goldberger-Treiman relations \cite{Goldberger:1958tr,Goldberger:1958vp}. In the chiral limit, this relation establishes a connection between the strong coupling of Goldstone bosons (here pions) to a hadron multiplet and the weak coupling of these hadrons. The coupling to external (weak) currents is best achieved by promoting the chiral symmetry to a local symmetry \cite{Scherer:2002tk}. 

We start with two baryon fields $B_1$ and $B_2$ of \emph{same} parity that behave oppositely with respect to
local chiral isospin transformations:
\begin{eqnarray}
  B_{1L}(x) & \to & U_L(x) \, B_{1L}(x) \,, \nonumber \\
  B_{1R}(x) & \to & U_R(x) \, B_{1R}(x) \,, \nonumber \\
  B_{2L}(x) & \to & U_R(x) \, B_{2L}(x) \,, \nonumber \\
  B_{2R}(x) & \to & U_L(x) \, B_{2R}(x) \,. 
  \label{eq:elemB12}
\end{eqnarray}
Here the label $L$ ($R$) for the baryon fields refers to left- (right-)handed fermions concerning the Lorentz group. Note the mirror assignment \cite{Detar:1988kn,Jido:2001nt} for $B_2$.  

To achieve local invariance of the Lagrangian, we introduce left- and right-handed external sources. Their transformation behavior is given by
\begin{eqnarray}
  l_\mu(x) & \to & U_L(x) \left( l_\mu(x) + i \partial_\mu \right) U_L^\dagger(x)  \,, \nonumber \\
  r_\mu(x) & \to & U_R(x) \left( r_\mu(x) + i \partial_\mu \right) U_R^\dagger(x)  \,.
  \label{eq:trafo-lr}  
\end{eqnarray}
Vector and axial-vector source terms are built in the following way:
\begin{eqnarray}
  v_\mu \equiv \frac12 \left(r_\mu + l_\mu \right) \,, \qquad a_\mu \equiv \frac12 \left( r_\mu - l_\mu \right)  \,. 
  \label{eq:defav}  
\end{eqnarray}
With the help of the external sources we can build chirally covariant derivatives. We will always use $D_\mu$, but it depends on the field
how $D_\mu$ is defined. Specifically
\begin{eqnarray}
  D_\mu B_{1L} \equiv \left(\partial_\mu - i l_\mu \right) B_{1L} \,, & \quad &  D_\mu B_{2R} \equiv \left(\partial_\mu - i l_\mu \right) B_{2R} \,, \nonumber \\ 
  D_\mu B_{1R} \equiv \left(\partial_\mu - i r_\mu \right) B_{1R} \,, & \quad &  D_\mu B_{2L} \equiv \left(\partial_\mu - i r_\mu \right) B_{2L} \,.
  \nonumber \\ 
  \label{eq:def-gaugecovder}  
\end{eqnarray}

Finally, we introduce the isospin matrix of scalar and pseudo-scalar meson fields,\footnote{Note that we use a combined isospin-spinor matrix in the main text to parameterize the mesons. The context will always reveal which matrix is used. Left- and right-handed fermion fields come with the isospin matrix specified in this appendix. Four-component fermion fields come with the meson matrix that contains a $\gamma_5$.} $M=\sigma + i \vec \tau \cdot \vec \pi$, which transforms according to
\begin{eqnarray}
  \label{eq:trafosigmapiM}
  M(x) \to U_L(x) M(x) U_R^\dagger(x) \,. 
\end{eqnarray}

We want to build chirally invariant fermion bilinears with one derivative at most. There are four groups distinct by chiral isospin and by
fermion chirality (Lorentz group property). These groups are
\begin{enumerate}
\setlength{\itemsep}{0pt}
\item $B_{1L}$ and $i \slashed D B_{2R}$,
\item $B_{2R}$ and $i \slashed D B_{1L}$,
\item $B_{1R}$ and $i \slashed D B_{2L}$,
\item $B_{2L}$ and $i \slashed D B_{1R}$. 
\end{enumerate}
Lorentz invariance prevents the combinations of group elements within one group and combinations of groups 1 and 4 and of groups 2 and 3.
Chiral isospin symmetry demands the use of a meson matrix $M$ or $M^\dagger$ when combining groups 1 and 3 or 2 and 4.
In total one can build $2^5=32$ terms. But 8 of them have two derivatives. Avoiding those, it leads to 24 terms.
As usual, total derivatives are ignored because they drop out in the action. This leads to the 20 terms given below.
Hermiticity and parity symmetry reduce them to 7 proper combinations and charge conjugation
symmetry makes the parameters real-valued. 
After normalizing the kinetic terms we find 
\begin{eqnarray}
  {\cal L}_{\rm ext} &=& \overline{B_{1L}} i \slashed D B_{1L} + \overline{B_{1R}} i \slashed D B_{1R}
  \nonumber \\ && {}
               + \overline{B_{2R}} i \slashed D B_{2R} + \overline{B_{2L}} i \slashed D B_{2L} \nonumber \\
           && {} -m_0 \big( \overline{B_{1L}} B_{2R} + \overline{B_{1R}} B_{2L} 
   \nonumber \\ && \phantom{mmmi} {}
           + \overline{B_{2R}} B_{1L} + \overline{B_{2L}} B_{1R} \big)
              \nonumber \\
           && {} -g_1 \left( \overline{B_{1L}} M B_{1R} + \overline{B_{1R}} M^\dagger B_{1L} \right)
            \nonumber \\&& {}
                            +g_2 \left( \overline{B_{2R}} M B_{2L} + \overline{B_{2L}} M^\dagger B_{2R} \right)  \nonumber \\
           && {} - h_1 \big( \overline{B_{1L}} M i \slashed D B_{2L} + \overline{B_{1R}} M^\dagger i \slashed D B_{2R}
           \nonumber \\ && \phantom{mmm} {}
              + \overline{i \slashed D B_{2L}} M^\dagger B_{1L} + \overline{i \slashed D B_{2R}} M B_{1R}\big)   \nonumber \\
           && {} - h_2 \big( \overline{B_{2R}} M i \slashed D B_{1R} + \overline{B_{2L}} M^\dagger i \slashed D B_{1L}
           \nonumber \\ && \phantom{mmm} {}
              + \overline{i \slashed D B_{1R}} M^\dagger B_{2R} + \overline{i \slashed D B_{1L}} M B_{2L}\big) \,.  \phantom{m}
  \label{eq:long-start}  
\end{eqnarray}
One can rewrite the fermions in a more compact form by using $B$ (and $\gamma_5 B$) instead of $B_L$ and $B_R$.
On the other hand, the coupling to mesons and to the
external currents needs then be rewritten into the explicit couplings to $\sigma$, $\pi$ and $a_\mu$, $v_\mu$, respectively.
We will do this in the end for the physical fields $N_+$ and $N_-$ that describe the nucleon and the $N^*$ resonance. 

It is worth to discuss the impact of the discrete symmetries in more detail.
Let us look at the $m_0$ terms in \eqref{eq:long-start}. Parity relates the first with the second term
(and the third with the fourth). Hermiticity relates the first with the third term. 
Charge conjugation maps a left-handed field, $B_L$, on the adjoint of a right-handed field, $B_R^\dagger$, and vice versa.
Thus it relates the first with the fourth term.
Parity alone would allow for a \emph{complex-valued} parameter $m_0$ in front of the first and second term.
Hermiticity demands that the parameter in front of the third and fourth term is $m_0^*$. Charge conjugation demands $m_0=m_0^*$.

As a second example we look at the $g_1$ terms. For the meson fields, the parity transformation is $M \to M^\dagger$.
Already parity symmetry demands the same coefficient in front of the two terms. Hermiticity makes $g_1$ real-valued. All this fits to charge
conjugation symmetry if the meson fields transform as $M \to M^T$. 

The analysis of the $m_0$ terms can be taken over to the $h_1$ and $h_2$ terms.
The appearance of $\gamma_\mu$ induces an additional sign change for charge conjugation. This is implicitly accounted for by extending the Dirac conjugation to the factor of $i$ in front of the derivative. 

If one analyzes the Lagrangian \eqref{eq:long-start} in the Wigner-Weyl phase of restored chiral isospin symmetry, the diagonalization procedure
leads to one positive and one negative eigenvalue for the mass parameters. To obtain the standard form of a free Lagrangian with two positive
mass parameters, one introduces
the new fields $N_{1L} \equiv B_{1L}$, $N_{1R} \equiv B_{1R}$, $N_{2L} \equiv -B_{2L}$, $N_{2R} \equiv +B_{2R}$; or in a more compact version:
\begin{eqnarray}
  \label{eq:translateBN}
  N_1 \equiv B_1 \,, \qquad N_2 \equiv \gamma_5 B_2  \,.
\end{eqnarray}
The new fields have now \emph{opposite} intrinsic parity relative to each other. Also charge conjugation looks slightly different because $N_{2L}$
is mapped on the adjoint of the negative of the right-handed field, i.e.\ on $-N_{2R}^\dagger$. In terms of these new fields the Lagrangian becomes 
\begin{eqnarray}
  {\cal L}_{\rm ext} &=& \overline{N_{1L}} i \slashed D N_{1L} + \overline{N_{1R}} i \slashed D N_{1R} 
  \nonumber \\ && {}
               + \overline{N_{2R}} i \slashed D N_{2R} + \overline{N_{2L}} i \slashed D N_{2L} \nonumber \\
           && {} -m_0 \big( \overline{N_{1L}} N_{2R} - \overline{N_{1R}} N_{2L} 
 \nonumber \\ && \phantom{mmm} {}          
           + \overline{N_{2R}} N_{1L} - \overline{N_{2L}} N_{1R} \big)
              \nonumber \\
           && {} -g_1 \left( \overline{N_{1L}} M N_{1R} + \overline{N_{1R}} M^\dagger N_{1L} \right)
            \nonumber \\&& {}
                            -g_2 \left( \overline{N_{2R}} M N_{2L} + \overline{N_{2L}} M^\dagger N_{2R} \right)  \nonumber \\
           && {} - h_1 \big( -\overline{N_{1L}} M i \slashed D N_{2L} + \overline{N_{1R}} M^\dagger i \slashed D N_{2R}
   \nonumber \\ && \phantom{mmmi} {}           
              - \overline{i \slashed D N_{2L}} M^\dagger N_{1L} + \overline{i \slashed D N_{2R}} M N_{1R}\big)   \nonumber \\
           && {} - h_2 \big( \overline{N_{2R}} M i \slashed D N_{1R} - \overline{N_{2L}} M^\dagger i \slashed D N_{1L}
   \nonumber \\ && \phantom{mmm} {}           
              + \overline{i \slashed D N_{1R}} M^\dagger N_{2R} - \overline{i \slashed D N_{1L}} M N_{2L}\big) \,. 
              \phantom{mm}
  \label{eq:long-start2}  
\end{eqnarray}
If we drop the coupling to external sources ($D \to \partial$) and introduce four-component spinors by using $\gamma_5$, we come to the Lagrangian \eqref{eq:Lagr-ext}. 

In this appendix, we will keep the coupling to external sources and also stick with the left- and right-handed fermion fields for some part of the analysis. 
Yet we put the free part of \eqref{eq:long-start2} into a more compact form, focusing now on the case of spontaneous symmetry breaking,
$M \to f_\pi$. We obtain 
\begin{eqnarray}
  {\cal L}_{\rm free} &=& \overline{N_{1}} i \slashed \partial N_{1} 
  + \overline{N_{2}} i \slashed \partial N_{2}  
                     -m_0 \left( \overline{N}_{1} \gamma_5 N_{2} - \overline{N}_{2} \gamma_5 N_{1} \right)  \nonumber \\
                      && {} -g_1 f_\pi \overline{N}_{1} N_{1} 
                         -g_2 f_\pi \overline{N}_{2} N_{2}   \nonumber \\
           && {} + 2 h f_\pi \left( \overline{N}_{1} \gamma_5 i \slashed \partial N_{2}
              +\overline{N}_{2} \gamma_5 i \slashed \partial N_{1}\right)  \,.
  \label{eq:long-start3}  
\end{eqnarray}
The parameters $h_1$ and $h_2$ enter the free Lagrangian only in the combination $h \equiv (h_1+h_2)/2$. 

Given the opposite intrinsic parities of $N_1$ and $N_2$, it is natural that $\overline{N}_{1} \gamma_5 N_{2}$ is the pertinent parity
symmetric combination. But a too naive
application of charge conjugation would map $\overline{N}_{1} \gamma_5 N_{2}$ on $+\overline{N}_{2} \gamma_5 N_{1}$. Hermiticity would
then demand that $m_0$ is purely imaginary. This is wrong. The correct charge conjugation operation is standard for $B_2$, but non-standard
for $N_2$, namely
\begin{eqnarray}
  \label{eq:charge-conj-special}
  \overline{N}_{1} \gamma_5 N_{2} = \overline{B}_1 B_2 \to + \overline{B}_2 B_1 = - \overline{N}_2 \gamma_5 N_1 \,.
\end{eqnarray}
We stress this explicitly because the common lore is that a pseudo-scalar bilinear is even with respect to charge conjugation. 

The remaining tasks are to diagonalize and rescale the free Lagrangian and to determine the coupling constants of the three-point vertices. One way
to achieve these goals is to start from the right- and left-handed fields in \eqref{eq:long-start} or \eqref{eq:long-start2} and introduce the
physical fermion fields via 
\begin{eqnarray}
\lefteqn{  \left(
  \begin{array}{c}
    N_{+L} \\ N_{+R} \\ N_{-L} \\ N_{-R} 
  \end{array}
  \right) =} \nonumber \\[0.5em] &&  \left(
  \begin{array}{cccc}
    \cos\theta_1 & 0 & \sin\theta_2 & 0 \\
    0 & \cos\theta_1 & 0 & \sin\theta_2 \\
    \sin\theta_1 & 0 & -\cos\theta_2 & 0 \\
    0 & -\sin\theta_1 & 0 & \cos\theta_2 
  \end{array}
  \right) \left(
  \begin{array}{c}
    B_{1L} \\ B_{1R} \\ B_{2L} \\ B_{2R} 
  \end{array}
  \right)  \,.
  \phantom{mm}
  \label{eq:large-rot}
\end{eqnarray}
The inverse relation is
\begin{eqnarray}
\lefteqn{  \left(
  \begin{array}{c}
    B_{1L} \\ B_{1R} \\ B_{2L} \\ B_{2R} 
  \end{array}
  \right) = \frac1{\cos(\theta_1-\theta_2)}  }  
  \nonumber \\ && \times
  \left(
  \begin{array}{cccc}
    \cos\theta_2 & 0 & \sin\theta_2 & 0 \\
    0 & \cos\theta_2 & 0 & -\sin\theta_2 \\
    \sin\theta_1 & 0 & -\cos\theta_1 & 0 \\
    0 & \sin\theta_1 & 0 & \cos\theta_1 
  \end{array}
  \right) 
  \left(
  \begin{array}{c}
    N_{+L} \\ N_{+R} \\ N_{-L} \\ N_{-R} 
  \end{array}
  \right)  \,.
  \phantom{mmm}
  \label{eq:large-rot-inv}
\end{eqnarray}
The corresponding relation for $N_{1,2}$ reads 
\begin{eqnarray}
\lefteqn{  \left(
  \begin{array}{c}
    N_{+L} \\ N_{+R} \\ N_{-L} \\ N_{-R} 
  \end{array}
  \right) =} 
  \nonumber \\ && \left(
  \begin{array}{cccc}
    \cos\theta_1 & 0 & -\sin\theta_2 & 0 \\
    0 & \cos\theta_1 & 0 & \sin\theta_2 \\
    \sin\theta_1 & 0 & \cos\theta_2 & 0 \\
    0 & -\sin\theta_1 & 0 & \cos\theta_2 
  \end{array}
  \right) \left(
  \begin{array}{c}
    N_{1L} \\ N_{1R} \\ N_{2L} \\ N_{2R} 
  \end{array}
  \right)  \,.
  \phantom{mm}
  \label{eq:large-rotN}
\end{eqnarray}
In compact notation this means
\begin{eqnarray}
  \label{eq:large-rotNusual}
  \left(
  \begin{array}{c}
    N_{+} \\ N_{-} 
  \end{array}
  \right) = \left(
  \begin{array}{cccc}
    \cos\theta_1 & \gamma_5 \sin\theta_2 \\
    - \gamma_5 \sin\theta_1 & \cos\theta_2 
  \end{array}
  \right) \left(
  \begin{array}{c}
    N_{1} \\ N_{2}
  \end{array}
  \right)  \,.
\end{eqnarray}

If one uses \eqref{eq:large-rot-inv}, then one can express the baryon masses and the diagonalization conditions in terms of the parameters of the
Lagrangian. One finds
\begin{eqnarray}
  m_+ &=& \frac1{\cos^2(\theta_1-\theta_2)} 
  \nonumber \\ & \times &
  \left( g_1 f_\pi \cos^2\theta_2 - g_2 f_\pi \sin^2\theta_1 + 2 m_0 \sin\theta_1 \cos\theta_2 \right) \,,
          \nonumber \\
  m_- &=& \frac1{\cos^2(\theta_1-\theta_2)} 
    \nonumber \\ & \times &
\left( -g_1 f_\pi \sin^2\theta_2 + g_2 f_\pi \cos^2\theta_1 + 2 m_0 \cos\theta_1 \sin\theta_2 \right) \,,
          \nonumber \\[0.5em]
  0 &=& g_1 f_\pi \sin(2\theta_2) + g_2 f_\pi \sin(2\theta_1) -2 m_0 \cos(\theta_1+\theta_2)  
  \nonumber \\ && 
  \label{eq:masses-fnc-param}  
\end{eqnarray}
together with
\begin{eqnarray}
  \label{eq:fixhfpi}
  2 h f_\pi = \sin(\theta_1-\theta_2)  \,. 
\end{eqnarray}
If one uses \eqref{eq:large-rot}, then one can express the parameters in terms of the physical baryon masses. In addition to \eqref{eq:fixhfpi},
this yields 
\begin{eqnarray}
  g_1 f_\pi &=& m_+ \cos^2 \theta_1 - m_- \sin^2\theta_1  \,, \nonumber \\
  g_2 f_\pi &=& m_- \cos^2 \theta_2 - m_+ \sin^2\theta_2  \,, \nonumber \\
  m_0 &=& m_+ \cos\theta_1 \sin\theta_2 +  m_- \cos\theta_2 \sin\theta_1 \,.
  \label{eq:param-fnc-masses}  
\end{eqnarray}
It has been cross-checked that the equations \eqref{eq:masses-fnc-param} and \eqref{eq:param-fnc-masses} are fully compatible
with each other.

If we keep only the external sources (and the vacuum expectation value of the meson-field matrix), the Lagrangian becomes 
\begin{eqnarray}
  {\cal L}_{\rm free + sources} &=& \overline{N_{+}} \, i \gamma^\mu \left(\partial_\mu - i v_\mu \right) N_+ - m_+ \overline{N_+} N_+ \nonumber \\
                                && {}+ \overline{N_{-}} \, i \gamma^\mu \left(\partial_\mu - i v_\mu \right) N_- - m_- \overline{N_-} N_- \nonumber \\
                                && {} + g_A^{++} \overline{N_{+}} \, a_\mu \gamma^\mu \gamma_5 N_+
                                \nonumber \\ && {}
                                   - g_A^{--} \overline{N_{-}} \, a_\mu \gamma^\mu \gamma_5 N_-  \nonumber \\
                                && {} - g_A^{+-} \left( \overline{N_{+}} \, a_\mu \gamma^\mu N_- + \overline{N_{-}} \, a_\mu \gamma^\mu N_+ \right)
                                \nonumber \\ &&
  \label{eq:free+sources}  
\end{eqnarray}
with the axial charges 
\begin{eqnarray}
  g_A^{++} &=& \frac{\cos(\theta_1+\theta_2) \cos(\theta_1-\theta_2)
  + 4 \Delta h f_\pi \sin\theta_1 \cos\theta_2 }{\cos^2(\theta_1-\theta_2)}   \,, \nonumber  \\ 
  g_A^{--} &=& \frac{\cos(\theta_1+\theta_2) \cos(\theta_1-\theta_2)
                   + 4 \Delta h f_\pi \cos\theta_1 \sin\theta_2 }{\cos^2(\theta_1-\theta_2)}   \,, \nonumber  \\
  g_A^{+-} &=& \frac{\sin(\theta_1+\theta_2) \cos(\theta_1-\theta_2) -2 \Delta h f_\pi \cos(\theta_1+\theta_2) }{\cos^2(\theta_1-\theta_2)} \,,
  \nonumber \\ &&
  \label{eq:gAetc}  
\end{eqnarray}
and $\Delta h \equiv (h_1-h_2)/2$. Note that we have used \eqref{eq:defav} to rewrite left- and right-handed source terms into vector and
axial-vector source terms. We see that the vector sources do not get renormalized. They relate to the unbroken vector subgroup. 

Now we turn to the meson-baryon three-point interactions. We will write down two versions, with and without using equations of motion.
For the latter case we obtain for the pions the Lagrangian
\begin{eqnarray}
  \lefteqn{{\cal L}_{\pi \, \neg \, \rm eom} = - \frac{g_1 \cos^2\theta_2 + g_2 \sin^2\theta_1}{\cos^2(\theta_1-\theta_2)} \,
                                       \overline{N_+} \, i \vec\tau \cdot \vec \pi \gamma_5 N_+} \nonumber \\[0.5em]
                                   && {}+\frac{g_1 \sin^2\theta_2 + g_2 \cos^2\theta_1}{\cos^2(\theta_1-\theta_2)} \,
                                       \overline{N_-} \, i \vec\tau \cdot \vec \pi \gamma_5 N_-  \nonumber \\[0.5em]
                                   && {}+ \frac{-g_1 \sin(2\theta_2) + g_2 \sin(2\theta_1)}{2\cos^2(\theta_1-\theta_2)}
        \nonumber \\ && \phantom{m} \times  
                                      \left( \overline{N_-} \, i \vec\tau \cdot \vec \pi N_+ - \overline{N_+} \, i \vec\tau \cdot \vec \pi N_- \right) \nonumber \\[0.5em]
                                   && {}+ \frac{2 \Delta h \sin\theta_1 \cos\theta_2}{\cos^2(\theta_1-\theta_2)} \,
                                      \overline{N_+} \, \vec\tau \cdot (\partial_\mu \vec \pi) \gamma^\mu \gamma_5 N_+ \nonumber \\[0.5em]
                                   && {}- \frac{2 \Delta h \cos\theta_1 \sin\theta_2}{\cos^2(\theta_1-\theta_2)} \,
                                      \overline{N_-} \, \vec\tau \cdot (\partial_\mu \vec \pi) \gamma^\mu \gamma_5 N_- \nonumber \\[0.5em]
                                   && {} + \frac{\Delta h \cos(\theta_1+\theta_2)}{\cos^2(\theta_1-\theta_2)}
        \nonumber \\ && \phantom{m} \times  
                                      \left(\overline{N_+} \, \vec\tau \cdot (\partial_\mu \vec \pi) \gamma^\mu N_-
                                      + \overline{N_-} \, \vec\tau \cdot (\partial_\mu \vec \pi) \gamma^\mu N_+ \right) 
     \nonumber \\[0.5em]  && {}
                 + \frac{h}{\cos(\theta_1-\theta_2)}  
    \Big( \overline{N_-} \vec\tau \cdot \vec \pi \slashed\partial N_+
    - \overline{\slashed \partial N_-} \vec\tau \cdot \vec \pi N_+
 \nonumber \\ && \hspace{7em} {}
                                       - \overline{N_+} \vec\tau \cdot \vec \pi \slashed\partial N_-
                                       + \overline{\slashed \partial N_+} \vec\tau \cdot \vec \pi N_- \Big)   \,.
\nonumber \\ &&
  \label{eq:reallylong}  
\end{eqnarray}
Note that the last term cannot be rewritten into a standard pseudo-vector interaction, i.e.\ we obtain more than just the standard
pseudo-scalar and pseudo-vector interactions when two different baryons are involved.

This pattern appears also for the three-point functions involving the $\sigma$ meson (we use $\tilde \sigma = \sigma - f_\pi$):
\begin{eqnarray}
  \lefteqn{{\cal L}_{\sigma \, \neg \, \rm eom} = \frac{-g_1 \cos^2\theta_2 + g_2 \sin^2\theta_1}{\cos^2(\theta_1-\theta_2)} \,
                                    \overline{N_+} \, \tilde\sigma N_+} \nonumber \\
                                   && {}+\frac{g_1 \sin^2\theta_2 - g_2 \cos^2\theta_1}{\cos^2(\theta_1-\theta_2)} \,
                                       \overline{N_-} \, \tilde\sigma  N_-  \nonumber \\
                                   && {}+ \frac{g_1 \sin(2\theta_2) + g_2 \sin(2\theta_1)}{2\cos^2(\theta_1-\theta_2)}
                                      \left( \overline{N_+} \, \tilde\sigma \gamma_5 N_- - \overline{N_-} \, \tilde\sigma \gamma_5 N_+ \right) \nonumber \\
                                   && {} + \frac{\Delta h}{\cos(\theta_1-\theta_2)}
                                      \left(\overline{N_+} \, i (\slashed \partial \tilde\sigma)  \gamma_5 N_-
                                      - \overline{N_-} \, i (\slashed \partial \tilde\sigma) \gamma_5 N_+ \right) \nonumber \\
                                   && {}- \frac{2 h \sin\theta_1 \cos\theta_2}{\cos^2(\theta_1-\theta_2)} \,
                                      \left( \overline{N_+} \, \tilde\sigma i \slashed\partial N_+ - \overline{\slashed\partial N_+} \, i \tilde\sigma N_+ \right)
                                      \nonumber \\
                                   && {}+ \frac{2 h \cos\theta_1 \sin\theta_2}{\cos^2(\theta_1-\theta_2)} \,
                                      \left( \overline{N_-} \, \tilde\sigma i \slashed\partial N_- - \overline{\slashed\partial N_-} \, i \tilde\sigma N_- \right)
                                      \nonumber \\
                                   && {}- \frac{h \cos(\theta_1+\theta_2)}{\cos^2(\theta_1-\theta_2)}  
\Big( \overline{N_-} i \tilde\sigma \slashed\partial \gamma_5 N_+
- \overline{\slashed \partial N_-} i \tilde\sigma \gamma_5 N_+
\nonumber \\ && \hspace{7em} {}
                                       + \overline{N_+} i \tilde\sigma \slashed\partial \gamma_5 N_-
                                       - \overline{\slashed \partial N_+} i \tilde\sigma \gamma_5 N_- \Big)  \,.
  \label{eq:reallylong-sigma}  
\end{eqnarray}

The equations of motion have the general structure
\begin{eqnarray}
  \label{eq:eom-use}
  (i \slashed \partial - m_{\pm}) N_{\pm} = J_{\pm} N_{\pm} 
\end{eqnarray}
where $J_{\pm}$ contains at least one field (meson or external source). Plugging the equation of motion back into the interaction Lagrangian
of three-point interactions yields new three-point interactions (with less many derivatives) and extra terms that contain $J_{\pm}$ together with
another meson field. Thus this procedure generates new four-point interactions. In other words, \emph{free} equations of motion can safely be
used to rewrite three-point interaction terms as long as one does not specify the four-point interactions.

In this spirit we can rewrite every derivative interaction into a pseudo-scalar interaction (and vice versa). Also the last
term in \eqref{eq:reallylong} can be simplified. This yields the Lagrangian for pion-baryon three-point interactions given in \eqref{eq:lagrpionbar-w-eom} of the main text.
Note that we have made use of \eqref{eq:fixhfpi} and \eqref{eq:param-fnc-masses}. We have also used \eqref{eq:gAetc} to show that we
have recovered the Goldberger-Treiman relations. 
Repeating this procedure for \eqref{eq:reallylong-sigma} leads to the Lagrangian for the $\sigma$ meson given in \eqref{eq:lagrsigmabar-w-eom}.

\section{Mean-field Lagrangian for general vacuum alignment}
\label{sec:MFL-VA}

Due to chiral symmetry we can generalize the free baryon Lagrangian of Subsection~\ref{subsec:KML} for an arbitrary spacetime constant meson matrix $M $, corresponding to a general alignment of the  $ O(4) $  vector $(\sigma,\vec\pi )$ on the vacuum manifold. This leads to the \emph{mean-field} Lagrangian 
\begin{align}
    \mathcal L_\mathrm{MF} &= \begin{pmatrix}
        \overline N_1, \overline N_2
    \end{pmatrix} 
    \nonumber \\[0.5em] & \times
    \begin{pmatrix}
        i\slashed{\partial} - g_1 M &  \gamma_5 ( 2 h M i\slashed{\partial} -m_0 ) \\
       \gamma_5 ( 2 h M^\dagger i\slashed{\partial} + m_0 )  & i\slashed{\partial} - g_2 M^\dagger 
    \end{pmatrix}
    \begin{pmatrix}
          N_1\\
          N_2
   \end{pmatrix}\, . \label{eq:MFlagr-chi-app}
\end{align}
This quadratic form defines in momentum space the Nambu-Gorkov-Dirac operator, $S^{-1}(p)  $, of the two-flavor parity-partner baryons $(N_1, N_2)$ for constant $\sigma $ and/or pion fields. Since $\gamma_0 S^{-1}(p) $ is hermitian, it can of course be diagonalized by a unitary transformation which here, for $h\not=0$, necessarily depends on the four momentum $p$ of the baryon, however, in addition to the meson matrix $M$ (and the constants $m_0$, $g_1$, $g_2$, $h$). In order to find the propagating normal modes corresponding to the physical mass and parity eigenstates, it turns out to be sufficient to apply a non-unitary but momentum-independent baryon-field transformation of a form, generalizing that in \eqref{eq:PhysFields}, 
\begin{equation}
   \begin{pmatrix}
          N_+\\
          N_-
   \end{pmatrix}
 = F
   \begin{pmatrix}
          N_1\\
          N_2
   \end{pmatrix}\, , \label{eq:PhysFieldsKM}
\end{equation}
with, again using the abbreviations $c_1= \cos\theta_1$, $s_1 =\sin\theta_1$ and $c_2 = \cos\theta_2$, $s_2 =\sin\theta_2$
\begin{equation}
   F
 =\begin{pmatrix}
          c_1           &  \gamma_5  s_2 \, U^\dagger \\
         -\gamma_5  s_1  &  c_2  \, U^\dagger
    \end{pmatrix}
  \, , \label{eq:defFieldTrafo-b}
\end{equation}
where the meson field dependence is encoded in the unitary matrix $U$, defined as in the main text by 
\begin{equation}
  U = \hat\sigma+i\gamma_5 \hat\pi\, , \;\; \hat\sigma = \sigma/\phi\, , \;\; \hat\pi = \vec\tau\vec\pi/\phi \, ,\;\; \phi^2=\sigma^2 +\vec\pi^2  \, ,  
\end{equation}
so that $M =\phi\, U$. This transformation is not measure preserving, but one has $\mathrm{det}\, F = \cos(\theta_1-\theta_2) $ which introduces a meson-field dependent Jacobian, as we will see below. Moreover, we will need
\begin{equation}
    F^\dagger F = \begin{pmatrix}
        1 & - \gamma_5 \sin(\theta_1-\theta_2) U^\dagger\\ 
       - \gamma_5 \sin(\theta_1-\theta_2) U & 1
    \end{pmatrix} \, , \label{eq:FdaggerF-app}
\end{equation}
and  
\begin{equation}
    \overline F = \gamma_0 F^\dagger \gamma_0 =\begin{pmatrix}
          c_1           &  \gamma_5  s_1  \\
         -\gamma_5  s_2 \, U^\dagger   &  c_2  \, U^\dagger
    \end{pmatrix} \,. 
\end{equation}
With these definitions, we express the mean-field Lagrangian in terms of the physical baryon fields as  
\begin{align}
    \mathcal L_\mathrm{MF} =&  \, \begin{pmatrix}
        \overline N_+, \overline N_-
    \end{pmatrix} 
    \begin{pmatrix}
        i\slashed{\partial} - m_+ U&  0 \\
       0  & i\slashed{\partial} - m_- U 
    \end{pmatrix}
    \begin{pmatrix}
          N_+\\
          N_-
   \end{pmatrix}  \nonumber \\ 
   =& \,  \begin{pmatrix}
        \overline N_1, \overline N_2
    \end{pmatrix} 
        i\slashed{\partial} \, F^\dagger F \,    \begin{pmatrix}
          N_1\\
          N_2
   \end{pmatrix} 
   \nonumber \\ & {}
   -   \begin{pmatrix}
        \overline N_1, \overline N_2
    \end{pmatrix} 
       \, \overline F \, \begin{pmatrix}
    m_+ U &  0 \\
       0  & m_- U
    \end{pmatrix} F \,    \begin{pmatrix}
          N_1\\
          N_2
   \end{pmatrix}
   \, .
\end{align}
Comparing this with \eqref{eq:MFlagr-chi-app}, using \eqref{eq:FdaggerF-app} we can first identify
\begin{equation}
    \sin(\theta_1-\theta_2) = 2 h\phi  \, ,
\end{equation} 
in agreement with Eq.~\eqref{eq:angles_1}. 
In addition, from
\begin{equation}
    \overline F \, \begin{pmatrix}
    m_+ U &  0 \\
       0  & m_- U
    \end{pmatrix} F \stackrel{!}{=}  \begin{pmatrix}
        g_1 M &  \gamma_5 m_0  \\
       -\gamma_5 m_0   & g_2 M^\dagger 
    \end{pmatrix} \, ,
\end{equation}
we obtain the relations
\begin{align}
    g_1 \phi &= \, c_1^2\, m_+ - s_1^2 \, m_- = \Delta m +\overline m  \, \cos 2\theta_1 \, , \nonumber \\
    g_2 \phi &= \, c_2^2\,  m_- - s_2^2\,  m_+ = - \Delta m + \overline m\, \cos 2 \theta_2 \, , \\
    m_0 &= \, c_1 s_2  \, m_+ + c_2 s_1\,  m_- 
    \nonumber \\ 
    &= \overline m \, \sin(\theta_1+\theta_2) - \Delta m \, \sin(\theta_1-\theta_2) \, . \nonumber
\end{align}
These are the generalizations, for general $M = \phi U$, of \eqref{eq:angles_2}, cf.~Sec.~\ref{sec:NGDH}.

The general form of the corresponding Dirac Hamiltonians
\begin{equation}
    H_\pm = \gamma_0 \vec\gamma\vec p +  \gamma_0 m_\pm U \, 
\end{equation}
is given in Eqs.~\eqref{eq:Hpgen}, \eqref{eq:Hmgen}.
These can be worked out explicitly, using Eq.~\eqref{eq:NGD}, to obtain,
\begin{align}
    H_+ = \, & \gamma_0 \vec\gamma\vec p +\frac{1}{1+2h\phi}
     \gamma_0 \Big( \frac{1}{2} (g_1-g_2) M -  m_0  U \Big) \cos^2\varphi \nonumber \\
      &  {}+
     \frac{1}{1-2h\phi} \gamma_0 
\Big(\frac{1}{2} (g_1-g_2) M +  m_0  U \Big) 
      \sin^2\varphi 
      \nonumber \\ & {}
      +  \frac{1}{\sqrt{1-4 h^2\phi^2 }} \, \gamma_0 (g_1+g_2 ) M  \cos\varphi\sin\varphi\, , \\
     H_- = \, & \gamma_0 \vec\gamma\vec p -\frac{1}{1+2h\phi}
     \gamma_0 \Big( \frac{1}{2} (g_1-g_2) M -  m_0  U \Big) \sin^2\varphi  \nonumber \\
      & {} -
     \frac{1}{1-2h\phi} \gamma_0 
\Big(\frac{1}{2} (g_1-g_2) M +  m_0  U \Big) 
      \cos^2\varphi 
      \nonumber \\ & {}
      +  \frac{1}{\sqrt{1-4 h^2\phi^2 }} \, \gamma_0 (g_1+g_2 )M  \cos\varphi\sin\varphi\, , 
\end{align}
where $\varphi = (\theta_1+ \theta_2)/2 +\pi/4 $ and $2 h\phi = \sin(\theta_1-\theta_2 )$ as in Sec.~\ref{sec:NGDH}.

As usual, they both satisfy  $H_\pm^2 = \vec p^{\, 2} + m_\pm^2 $, so that we can read off the mass eigenvalues for general vacuum alignment as
\begin{align}
    m_+ = \, &     \frac{1}{1+2h\phi}  \Big( \frac{1}{2} (g_1-g_2) \phi  -  m_0  \Big) \cos^2\varphi   
    \nonumber \\ &  {}
    + \frac{1}{1-2h\phi} 
\Big(\frac{1}{2} (g_1-g_2) \phi  +  m_0  \Big) 
      \sin^2\varphi \label{eq:mp0}\\
     & {}+  \frac{1}{\sqrt{1-4 h^2\phi^2 }} \, \frac{1}{2} (g_1+g_2 )\phi \sin 2\varphi \, , \nonumber \\
    m_- = \, & -\frac{1}{1+2h\phi}  \Big( \frac{1}{2} (g_1-g_2) \phi  -  m_0  \Big) \sin^2\varphi   
    \nonumber \\ & {}
    - \frac{1}{1-2h\phi} 
\Big(\frac{1}{2} (g_1-g_2) \phi  +  m_0  \Big) 
      \cos^2\varphi \label{eq:mm0}\\
     & {} +  \frac{1}{\sqrt{1-4 h^2\phi^2 }} \, \frac{1}{2} (g_1+g_2 )\phi \sin 2\varphi\,. \nonumber
\end{align}
Without kinetic mixing, i.e.~for $h=0$ and with $\varphi = \theta +\pi/4$ this reduces to the masses of nucleon and $N^* $ in Eq.~\eqref{eq:mass_pm}.

Using Eq.~\eqref{eq:varphi} we can furthermore eliminate $m_0 $ from Eqs.~\eqref{eq:mp0} and \eqref{eq:mm0}, writing
\begin{align}
    m_+ &=\frac{1}{2}(g_1-g_2) \phi + \frac{1}{2}(g_1+g_2)\phi \, \frac{1-2h \phi \cos 2\varphi}{\sqrt{1-4h^2\phi^2}\, \sin 2\varphi} \,,\\
    m_- &= - \frac{1}{2}(g_1-g_2) \phi + \frac{1}{2}(g_1+g_2)\phi\,  \frac{1+ 2h \phi \cos 2\varphi}{\sqrt{1-4h^2\phi^2}\, \sin 2\varphi } \,,
    \label{eq:Hamiltonian_massesA}
\end{align}
yielding Eqs.~\eqref{eq:Hamiltonian_massp}, \eqref{eq:Hamiltonian_massm} in the main text.

\section{The $\sigma$ meson as a broad resonance of pion-pion scattering}
\label{sec:sigma-broad}

The PDM provides predictions for the coupling of the $\sigma$ meson to baryons. In principle, the prediction for the $\sigma$-$N^*$-$N$ interaction strength can be compared to the partial decay width of $N(1535) \to \sigma N$. In general, we stick to tree-level calculations in the present work. However, the $\sigma$ meson is a very broad resonance \cite{ParticleDataGroup:2024cfk} of pion-pion scattering \cite{Pelaez:2015qba}. Approximating it by a stable state would yield very misleading results. Also on the experimental side, the $\sigma$ meson is extracted from events with two pions in the final state \cite{CBELSATAPS:2015kka,Sarantsev:2025lik}. This requires the introduction of a realistic spectral distribution for the $\sigma$ meson. This is the main aspect of this section. However, we establish first the decay formulae. 

We start with the three-point interaction between a $\sigma$ meson, a nucleon and a $J^P = \frac12^-$ resonance,
\begin{eqnarray}
  {\cal L} = g_{\sigma NN^*} \left(\overline N \tilde \sigma \gamma_5 N^* - \overline N^* \tilde \sigma \gamma_5 N  \right) \,.
  \label{eq:genNsigmaNstar}
\end{eqnarray}
The Feynman matrix element for the decay $N^* \to N \sigma$ is given by
\begin{eqnarray}
  {\cal M} = g_{\sigma NN^*} \, \bar u_N(p') \gamma_5 u_{N^*}(p)  \,.
  \label{eq:MdecNNstsigma}
\end{eqnarray}
This leads to the spin-averaged squared matrix element
\begin{eqnarray}
  \langle \vert {\cal M} \vert^2 \rangle 
  &=& -\frac12 \vert g_{\sigma NN^*} \vert^2 {\rm Tr}\left[ (\slashed p' + m_N) \gamma_5 (\slashed p + m_{N^*}) \gamma_5 \right]
  \nonumber \\
  &=& \vert g_{\sigma NN^*} \vert^2 \left[ (m_{N^*}-m_N)^2- m_\sigma^2 \right]  \,.
  \label{eq:MdecNNstsigma-sq}  
\end{eqnarray}
One reason why we write this down in detail is to stress that this relativistic version is {\em not} fully compatible with the following guess inspired by a non-relativistic treatment: 
\begin{eqnarray}
    && \langle \vert {\cal M} \vert^2 \rangle  \stackrel{?}  {\sim}  \vert \vec p_\sigma \vert^2 
    \nonumber \\
    &&= \frac1{4 m_{N^*}^2} 
    \left[ (m_{N^*}-m_N)^2- m_\sigma^2 \right] 
    \left[ (m_{N^*}+m_N)^2- m_\sigma^2 \right] \,.
    \nonumber \\ &&
\end{eqnarray}
The extra factor $(m_{N^*}+m_N)^2- m_\sigma^2$, which depends on the $\sigma$ mass, does not show up in \eqref{eq:MdecNNstsigma-sq}. This is a serious issue because the $\sigma$ meson is a very broad resonance. This makes it necessary to introduce a spectral distribution where the $\sigma$ mass becomes a variable. For such a case, the partial decay width is given by 
\begin{eqnarray}
 \lefteqn{  \Gamma_{N^* \to (\pi\pi)_{\sigma} N} 
  =   \vert g_{\sigma N N^*} \vert^2 \int\limits_{4 m_\pi^2}^{(m_{N^*}-m_N)^2} dm_{12}^2 \, \frac1{\cal N} {\cal A}(m_{12}^2) }
  \nonumber \\ && \phantom{mm} \times
  \frac1{8 \pi} \frac{\lambda^{1/2}(m^2_{N^*},m_N^2,m_{12}^2)}{2 m^3_{N^*}} 
\left[ (m_{N^*}-m_N)^2- m_{12}^2 \right]  \,.
\nonumber \\ &&
  \label{eq:width-broad-sigma}
\end{eqnarray}
We have introduced the 
K\"all\'en function
\begin{eqnarray}
  \label{eq:kallenfunc}
\lambda(a,b,c) \equiv a^2+b^2+c^2-2(ab+bc+ac) \,,
\end{eqnarray}
here leading to
\begin{eqnarray}
\lambda(m^2_{N^*},m_N^2,m_{12}^2) &=& \left[ (m_{N^*}-m_N)^2- m_{12}^2 \right] 
\nonumber \\ && \times
\left[ (m_{N^*}+m_N)^2- m_{12}^2 \right] \,.
\phantom{m}
  \label{eq:kallen-var}
\end{eqnarray}
This function serves to specify the momentum of the nucleon and the sigma meson in the rest frame of the decaying resonance
($\vert \vec p \vert = \lambda^{1/2}/(2 m_{N^*})$). 

The spectral distribution of the $\sigma$ meson is denoted by ${\cal A}$. The quantity ${\cal N}$ denotes the normalization of this spectral distribution:
\begin{eqnarray}
  \label{eq:normal-spec}
  {\cal N} \equiv \int\limits_{4 m_\pi^2}^\infty dm_{12}^2 \, {\cal A}(m_{12}^2)  \,.
\end{eqnarray}

Note that in \eqref{eq:width-broad-sigma} one finds the invariant two-pion mass $m_{12}$ instead of an onshell $\sigma$ mass in the phase space factor 
$\lambda^{1/2}$ and in the $(m_{N^*}-m_N)^2-m_{12}^2$ factor of the squared matrix element. 
What makes the introduction of a spectral distribution indispensable is the kinematical situation that we face: 
A stable sigma meson would grossly underestimate the available phase space that starts at $m_N + 2 m_\pi$ and not only at $m_N+ m_\sigma$. For narrow resonances, the correct threshold is of minor importance. But for broad resonances, the correct threshold has an important quantitative impact. 

We need to determine the spectral distribution of the $\sigma$ meson. We introduce the propagator
\begin{eqnarray}
  \label{eq:sigma-prop-with-Pi}
  D_\sigma(s) = \frac1{s-\mbare^2 - \Pi(s)} 
\end{eqnarray}
with the self-energy $\Pi(s)$ and define
\begin{eqnarray}
  \label{eq:def-spec}
  {\cal A} \equiv -\frac1{\pi} {\rm Im} D_\sigma \,.
\end{eqnarray}
The quantity $\mbare$ denotes the bare mass of the $\sigma$ meson. By the interactions with the pions the bare, stable $\sigma$ becomes a resonance with a pole of the scattering amplitude on the second Riemann sheet of two-pion scattering. 

Let us briefly review the most important aspects of the $\sigma$ meson and the scattering of two pions in an s-wave with isospin 0.
The $\sigma$ meson aka $f_0(500)$ is the lightest hadronic resonance. Its pole on the second Riemann sheet lies far away from the real axis \cite{Pelaez:2015qba}. This makes its description by a Breit-Wigner function problematic. At least, one should include an energy dependent real part of a self-energy and not only a width. The $\sigma$ meson is an elastic resonance of pion-pion scattering in the s-wave of isospin zero. Thus we should look at the properties of the $\sigma$ meson together with the properties of the pion-pion scattering amplitude. 

The first idea would be simply to approximate the self-energy of the $\sigma$ meson by a one-loop expression with two pion propagators. Such a loop expression is given by
\begin{equation}
  \label{eq:defcapital-I}
  I(q^2) \equiv -i \int \frac{d^4l}{(2\pi)^4} \frac{1}{l^2-m_\pi^2+ i \epsilon} \frac{1}{(q-l)^2-m_\pi^2+ i \epsilon}  \,.
\end{equation}
Its renormalization is discussed below. Our first guess is then $\Pi \stackrel{?}{=} -\frac32 \ourg^2 I$ where $\ourg$ denotes the coupling of the $\sigma$ meson to charged pions. The self-energy accounts for pion loops with charged and neutral pions. This leads to the coefficient $3/2$.

One could adjust the parameters (bare $\sigma$ mass $\mbare$ and $\sigma$-pion coupling $\ourg$) such that the pole on the second Riemann sheet agrees with the phenomenological value given in \cite{Pelaez:2015qba}. However, we have checked that this construction leads to an additional pole on the real axis just below the two-pion threshold, a very unphysical artifact. We do not show the results of this approach. A similar line of reasoning against the use of simple one-loop self-energies has been provided in \cite{Heuser:2024biq}. 

The conclusion is that we need a better account of the sub-threshold region of pion-pion scattering for the isospin-zero $s$-wave. An important aspect is here the appearance of an Adler zero (see \cite{GarciaMartin:2011cn} and references therein). To account for this significant feature we start from an approximation to the pion-pion scattering amplitude instead of an approximation to the $\sigma$ self-energy. In other words, we start from a two-body quantity instead of a one-body quantity. We want to account for a resonance and for the Adler zero. The key is the use of an extra contact interaction
(a natural ingredient in effective field theories) besides the resonance propagator. In the present work, we describe a formalism for the $\sigma$ meson and pion scattering in the $s$-wave that has been developed in \cite{Leupold:2009nv}, originally for the $\rho$ meson and the $p$-wave. 

Let us provide some basic definitions. One subtraction is required to make the loop function $I$ finite (in four dimensions). One obtains (see e.g.\ \cite{Leupold:2009nv}):
\begin{eqnarray}
  \label{eq:pi-diff}
  I(q^2) - I(0) = J(q^2)
\end{eqnarray}
where on the real axis\footnote{Strictly speaking $J$ has a right-hand cut along the real axis starting at the two-pion threshold. The expression for $J$ is valid above the cut, i.e.\ for $s + i\epsilon$.} $J$ is given by 
\begin{eqnarray}
  \label{eq:Jint}
  J(s) = \frac1{16 \pi^2} \left( 2 + \sigma(s) \log \frac{\sigma(s)-1}{\sigma(s) +1} \right)
\end{eqnarray}
with the velocity of the pions in the two-particle rest frame
\begin{eqnarray}
  \label{eq:defsigma}
  \sigma(s) \equiv \sqrt{1-\frac{4m_\pi^2}{s}}  \,.
\end{eqnarray}
For later use we provide
\begin{eqnarray}
  \label{eq:imag}
  {\rm Im}I(s) = {\rm Im}J(s) =  \frac{\sigma(s)}{16 \pi}
\end{eqnarray}
which holds along the real axis for $s \ge 4 m_\pi^2$. Below the two-pion threshold the $\sigma$ meson has no width (in vacuum). It is straightforward to extend the loop functions $I$ and $J$ to the two Riemann sheets that emerge from the cut caused by elastic rescattering of the two pions \cite{Moussallam:2011zg}. 

Now we turn to the two-body quantity, the scattering amplitude $t$, as obtained from a Bethe-Salpeter equation (ignoring renormalization for a moment)
\begin{eqnarray}
    t(s) = k(s) + 16 \pi k(s) I(s) t(s)
\end{eqnarray}
which one can rephrase as
\begin{eqnarray}
  \label{eq:BS-SL}
  t^{-1}(s) = k^{-1}(s) - 16\pi I(s)
\end{eqnarray}
with a scattering kernel $k(s)$ that has no right-hand cut. This relation ensures the optical theorem $\mathrm{Im}\, t = \sigma \, t \, t^*$, 
which can be recast into $\mathrm{Im}\,t^{-1} = - \sigma $.
A graphical representation can be found in Figure 1 of \cite{Leupold:2009nv}.

We approximate the scattering kernel by the sum of a contact term and a (tree-level) resonance.
The contact term is supposed to serve as an
approximation for the left-hand cut structures (in particular $\rho$ and $\sigma$ exchange). We demand two properties for our scattering
amplitude to hold. It should have a resonance pole on the second Riemann sheet, as discussed before. In addition, the scattering kernel and
therefore the scattering amplitude should have an Adler zero \cite{GarciaMartin:2011cn} at the appropriate location.
We deem this as the most important chiral constraint for the isoscalar $s$-wave amplitude.

The scattering kernel is approximated by 
\begin{eqnarray}
  k(s) = -\frac3{32\pi} \ourg^2 \frac1{s-\mbare^2} + \frac3{32\pi} \ourg^2 \frac1{s_A-\mbare^2}
  \label{eq:k-tree}  
\end{eqnarray}
and we renormalize at the Adler zero $s_A$. Thus, the appropriate proper version of (\ref{eq:BS-SL}) is
\begin{eqnarray}
  \label{eq:BS-SL-true}
  t^{-1}(s) = k^{-1}(s) - 16\pi \left( J(s) - J(s_A) \right)  \,.
\end{eqnarray}
We recall that $\mbare$ denotes the bare mass of the $\sigma$ meson and $\ourg$ the coupling of the $\sigma$ to two charged pions. 

In leading order chiral perturbation theory \cite{Gasser:1983yg}, the Adler zero in the scalar isoscalar channel sits at 
\begin{eqnarray}
  \label{eq:Adler-zero}
  s_A^{\rm LO} = \frac12 m_\pi^2  \approx (98 \, {\rm MeV})^2 \,.
\end{eqnarray}
where we have used an isospin  averaged value of $m_\pi \approx 138 \,$MeV for the pion mass \cite{ParticleDataGroup:2024cfk}. 
According to the dispersive analysis \cite{GarciaMartin:2011cn} of pion-pion scattering, the Adler zero is located at
\begin{eqnarray}
  \label{eq:Adler-zero-disp}
  s_A \approx (84 \, {\rm MeV})^2 
\end{eqnarray}
but with a relatively large uncertainty that covers the leading-order result. We have checked that both choices lead practically to the very same
propagator of the $\sigma$ meson. Therefore we use (\ref{eq:Adler-zero}) in the following. For most of our calculations we will also use the isospin averaged pion mass. Yet when comparing our pion phase shift to the results from \cite{GarciaMartin:2011cn} we also readjust the pion mass to the choice of \cite{GarciaMartin:2011cn}, $m_\pi \approx 139.57\,$MeV, such that the phase shift can be meaningfully compared also close to threshold. Except there, the differences are minute. 

We adjust our parameters $\ourg$ and $\mbare$ such that the pole of the scattering amplitude lies (on the second Riemann sheet) at the phenomenologically determined location \cite{Pelaez:2015qba}
\begin{eqnarray}
  \label{eq:pole-numerical}
  s_{\rm pole} = \left((0.45 \pm 0.02) - i (0.28 \pm 0.01) \right)^2 \, \mbox{GeV}^2.
  \phantom{mm}
\end{eqnarray}
We obtain
\begin{eqnarray}
  \label{eq:solvembare}
  \mbare^2 \approx 0.67 \, \mbox{GeV}^2 \qquad \Rightarrow \qquad \mbare \approx 0.82 \, \mbox{GeV.}
  \phantom{m}
\end{eqnarray}
We note in passing that this is a quite interesting result for the tree-level mass of the $\sigma$ meson. Based on algebraically realized chiral symmetry, 
Weinberg suggested \cite{Weinberg:1969hw} that the mass of the $\sigma$ meson should coincide with the mass of the $\rho$ meson
(in the chiral limit).
The result (\ref{eq:solvembare}) is close to the mass of the $\rho$ meson.

\begin{figure}[t]
\centering
    \includegraphics[width=0.95\linewidth]{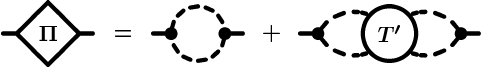}
    \caption{Calculation of the $\sigma$ self-energy beyond one-loop by allowing the pions to rescatter.
      Figure taken from \cite{Leupold:2009nv}.}
    	\label{fig:selfen-gr}
\end{figure}

For the coupling of the $\sigma$ meson to two charged pions we find
\begin{eqnarray}
  \label{eq:val-g2-impr}
  \ourg^2 \approx 33.6 \, \mbox{GeV}^{2}.
\end{eqnarray}

With all parameters determined, we can turn to an improved version of the self-energy of the $\sigma$ meson, following \cite{Leupold:2009nv}. For convenience we introduce the scattering amplitude $t'$ that would emerge
from a pure contact interaction without the $\sigma$ resonance. It is given by 
\begin{eqnarray}
  \label{eq:t-prime-sol}
  t'(s) = \frac3{32\pi} \ourg^2 \frac1{s_A-\mbare^2-\frac32 \ourg^2 \left( J(s) - J(s_A) \right)}  \,.
  \phantom{m}
\end{eqnarray}
A graphical representation can be found in Fig.~2 of \cite{Leupold:2009nv}.

\begin{figure*}[t]
\centering    
\includegraphics[width=0.475\textwidth]{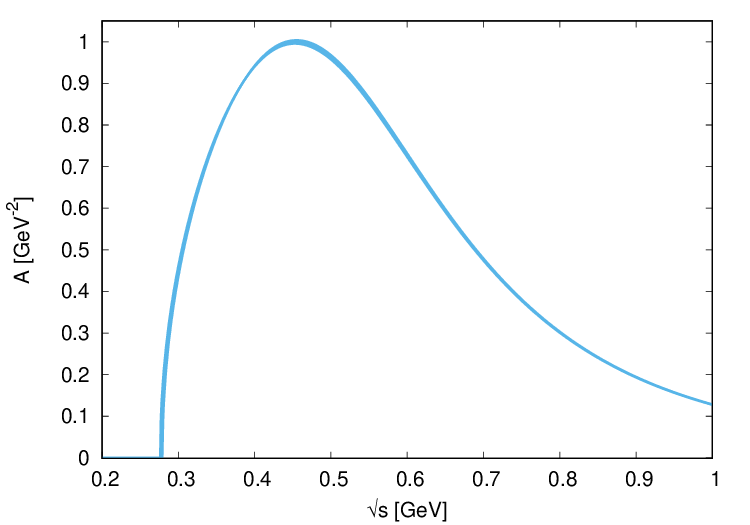}
\hspace{.4cm}
\parbox[b]{0.42\textwidth}{\includegraphics[width=0.42\textwidth]{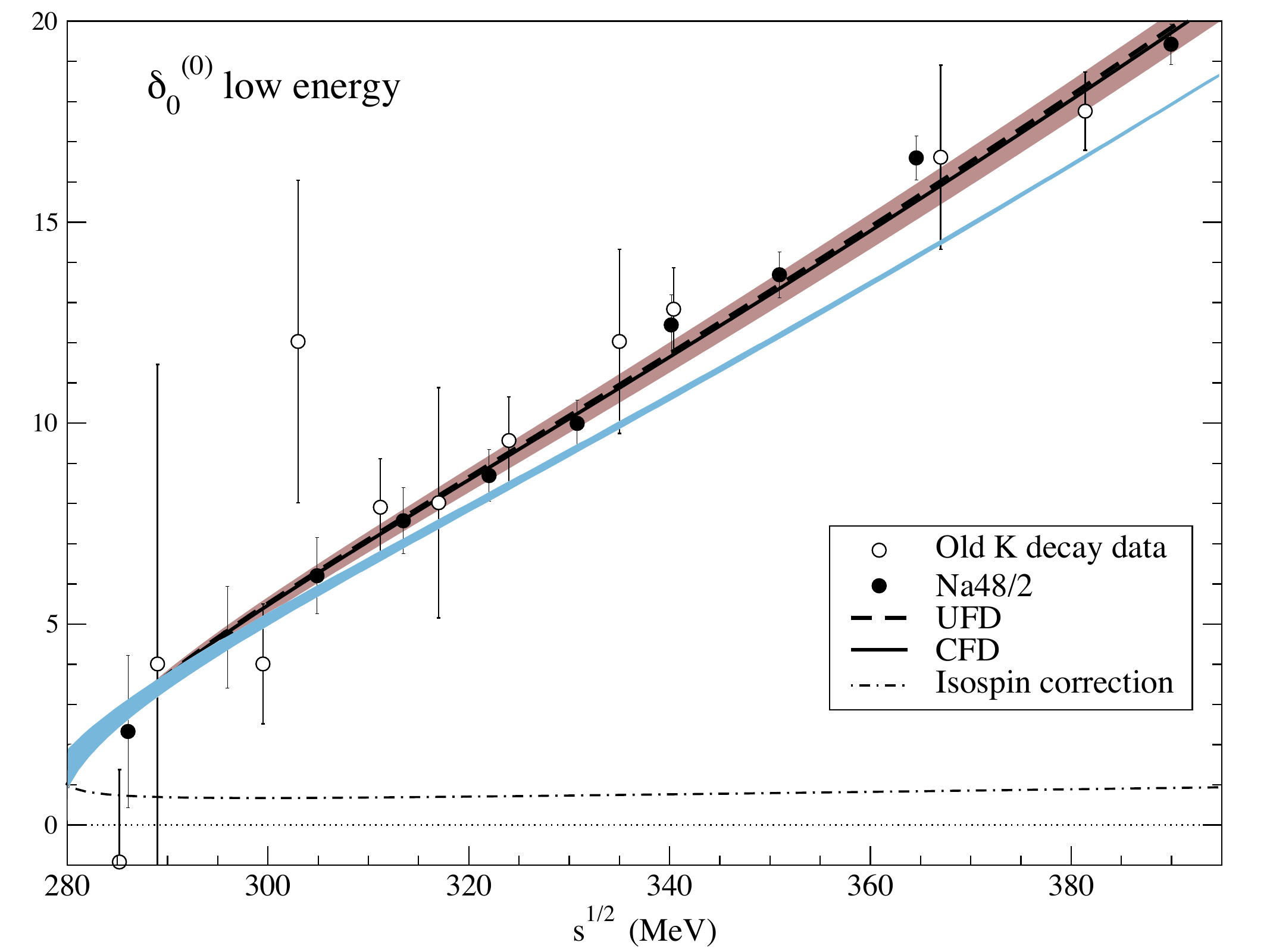}
\vspace{-6pt}
}
\caption{{\it Left:}  Spectral distribution ${\cal A}$ for the $\sigma$ meson as a function of the energy $\sqrt{s}$. {\it Right:} Comparison of our result for the pion phase shift (blue) to the dispersive analysis of \cite{GarciaMartin:2011cn}. The underlying figure is taken from \cite{GarciaMartin:2011cn}. The width of the blue lines (our results) in the two panels is obtained by varying the pion mass between the isospin average (our original choice for the calculations) and the charged pion mass (the choice of \cite{GarciaMartin:2011cn}).
        } 
    	\label{fig:spec-sigma-impr}
        \vspace{-6pt}
\end{figure*}

We propose that the complete solution of the Bethe-Salpeter equation (\ref{eq:BS-SL-true}) can be
decomposed as 
\begin{eqnarray}
  t(s) &=& t'(s)- \frac32 \ourg^2 t'(s) \left( J(s) - J(s_A) \right) D_\sigma(s) \nonumber \\
  && {}  - \frac32 \ourg^2 D_\sigma(s) \left( J(s) - J(s_A) \right) t'
    - \frac3{32 \pi} \ourg^2 D_\sigma(s) 
 \nonumber \\
  && {}   
     - \frac32 \ourg^2 t'(s) \left( J(s) - J(s_A) \right) D_\sigma(s) 
     \nonumber \\ && \phantom{m} \times
     16 \pi \left( J(s) - J(s_A) \right) t'(s)   \label{eq:decompose-t-via-t-prime}  
\end{eqnarray}
i.e.\ we consider all the possibilities that the incoming or outgoing pions interact with or without forming the $\sigma$ resonance.
Of course, it is not possible to
form the resonance twice. That would spoil the resummation of one-particle reducible diagrams.
A graphical representation can be found in Figure 4 of \cite{Leupold:2009nv}.

By comparison of (\ref{eq:BS-SL-true}) and
(\ref{eq:decompose-t-via-t-prime}) we find \cite{Leupold:2009nv}
\begin{eqnarray}
  \label{eq:self-improv-BS}
  \Pi(s) &=& -\frac32 \ourg^2 \left( J(s) - J(s_A) \right) 
  \nonumber \\ && \times
  \frac{s_A-\mbare^2}{s_A-\mbare^2-\frac32 \ourg^2 \left( J(s) - J(s_A) \right)}  
\end{eqnarray}
where we have used the Dyson-Schwinger equation \eqref{eq:sigma-prop-with-Pi}. A graphical representation is provided in Figure \ref{fig:selfen-gr}.

Note that the pion-loop terms $\sim J$ in the denominator of \eqref{eq:self-improv-BS} describe the rescattering of pions. Dropping those would bring us back to the standard one-loop expression for the self-energy (renormalized at the Adler zero). But the inclusion of the rescattering is important to improve on the properties of the broad $\sigma$ resonance.

The propagator \eqref{eq:sigma-prop-with-Pi} obtained from this improved self-energy does not have a single-particle pole close to the two-pion threshold, a substantial improvement over a simple one-loop self-energy. However, one finds a pole in the deep spacelike region at about $-2\,$GeV$^2$, i.e.\ far away from the low-energy regime for which this model has been developed. The only consequence of this extra strength is the fact that the spectral distribution is not properly normalized. With the definition \eqref{eq:normal-spec} we obtain ${\cal N} \approx 0.56$.
We have obtained the spectral distribution from \eqref{eq:def-spec} and \eqref{eq:sigma-prop-with-Pi}. It is plotted in Figure \ref{fig:spec-sigma-impr}, left panel. One observes a broad peak at around $\sqrt{s_{\rm peak}} = 0.46 \,$GeV
which agrees nicely with the real part of the pole position provided in \eqref{eq:pole-numerical}. 

We also note that the pion phase shift for the scalar-isoscalar channel  \cite{Pelaez:2015qba,GarciaMartin:2011cn} looks very decent. The relation between the scattering amplitude $t$ and the phase shift $\delta$ is given by
\vspace{-12pt}
\begin{eqnarray}
t = \frac{e^{i\delta} \sin\delta}{\sigma}  \,.
    \label{eq:phaseshift-scatt}
\end{eqnarray}
Our phase shift is shown in Figure \ref{fig:spec-sigma-impr}, right panel, where we also compare to the results of \cite{GarciaMartin:2011cn}. Keeping in mind the simplicity of our model, we obtain a very reasonable result. Note that for higher energies (not shown here) the phenomenological phase shift shows the influence of the $K \bar K$ inelasticity with the appearance of the $f_0(980)$ \cite{GarciaMartin:2011cn}. Also the four-pion channel ($\sigma \sigma$) might play a role. Such effects are not covered by our approach, but we do not expect that this matters much for the study of the baryon decays $N^* \to N (\pi \pi)_\sigma$.

With the self-energy \eqref{eq:self-improv-BS} entering the propagator \eqref{eq:sigma-prop-with-Pi} and the spectral distribution \eqref{eq:def-spec} we are equipped to deduce the $\sigma$-baryon coupling strength from \eqref{eq:width-broad-sigma}.

\bibliographystyle{apsrev4-2}
\bibliography{literature.bib}

\end{document}